\documentclass[11pt,preprint]{emulateapj}
\usepackage{epsfig,natbib}
\usepackage{float}
\usepackage{soul}
\usepackage{graphicx} 

\usepackage[usenames,dvipsnames]{color}

\def\bluetext#1{{\color{blue}#1}}

\usepackage[normalem]{ulem}             
\newcommand\redsout{\bgroup\markoverwith{\textcolor{red}{\rule[0.5ex]{2pt}{0.4pt}}}\ULon}
\newcommand\bluesout{\bgroup\markoverwith{\textcolor{blue}{\rule[0.5ex]{2pt}{0.6pt}}}\ULon}
\def\bluesout#1{}
\newcommand\greensout{\bgroup\markoverwith{\textcolor{Green}{\rule[0.5ex]{2pt}{0.6pt}}}\ULon}
\def\greensout#1{}
\newcommand\magsout{\bgroup\markoverwith{\textcolor{Magenta}{\rule[0.5ex]{2pt}{0.4pt}}}\ULon}
\newcommand\bricksout{\bgroup\markoverwith{\textcolor{Brickred}{\rule[0.5ex]{2pt}{0.4pt}}}\ULon}

\newcommand\reduline{\bgroup\markoverwith{\lower 3.5 pt\hbox{\textcolor{red}{\rule[0.5ex]{2pt}{0.4pt}}}}\ULon}
\newcommand\blueuline{\bgroup\markoverwith{\lower 3.5 pt\textcolor{blue}{\rule[0.5ex]{2pt}{0.4pt}}}\ULon}
\newcommand\greenuline{\bgroup\markoverwith{\lower 3.5 pt\textcolor{Green}{\rule[0.5ex]{2pt}{0.4pt}}}\ULon}
\newcommand\maguline{\bgroup\markoverwith{\lower 3.5 pt\textcolor{Magenta}{\rule[0.5ex]{2pt}{0.4pt}}}\ULon}
\newcommand\brickuline{\bgroup\markoverwith{\lower 3.5 pt\textcolor{Brickred}{\rule[0.5ex]{2pt}{0.4pt}}}\ULon}

\shortauthors{Liu, Eracleous, \& Halpern}

\def\Msol{\ifmmode{\rm M}_{\mathord\odot}\else M$_{\mathord\odot}$\fi}

\def\ls{\lower 2pt \hbox{$\scriptscriptstyle \buildrel<\over\sim$}} 
\def\gs{\lower 2pt \hbox{$\scriptscriptstyle \buildrel>\over\sim$}} 

\def\aj{\rm{AJ}}                   
\def\apj{\rm {ApJ}}                
\def\apjl{\rm{ApJ}}                
\def\apjs{\rm{ApJS}}               
\def\apss{\rm{Ap\&SS}}             
\def\aap{\rm{A\&A}}                
\def\mnras{\rm{MNRAS}}             
\def\nat{\rm{Nature}}              
\def\pasj{\rm{PASJ}}    	   
\def\nar{\rm{New Astr. Rev.}}

\begin{document}
\title{A Radial Velocity Test for Supermassive Black Hole Binaries as an Explanation for Broad, Double-Peaked Emission Lines in Active Galactic Nuclei}
\author{Jia Liu\altaffilmark{1}, Michael Eracleous\altaffilmark{2}, and Jules P. Halpern\altaffilmark{1}}

\altaffiltext{1}{Astronomy Department, Columbia
  University, 550 West 120th Street, New York, NY 10027, U.S.A.}
\altaffiltext{2}{Department of Astronomy and Institute for Gravitation
  and The Cosmos, The Pennsylvania State University,
  525 Davey Lab, University Park, PA 16802, U.S.A.}


\begin{abstract}

One of the proposed explanations for the broad, double-peaked Balmer emission lines observed in the spectra of some active galactic nuclei (AGNs) is that they are associated with sub-parsec supermassive black hole (SMBH) binaries. Here, we test the binary broad-line region hypothesis through several decades of monitoring of the velocity structure of double-peaked H$\alpha$ emission lines in 13 low-redshift, mostly radio-loud AGNs. This is a much larger set of objects compared to an earlier test by \cite{Eracleous1997} and we use much longer time series for the three objects studied in that paper. Although systematic changes in radial velocity can be traced in many of their lines, they are demonstrably not like those of a spectroscopic binary in a circular orbit. Any spectroscopic binary period must therefore be much longer than the span of the monitoring (assuming a circular orbit), which in turn would require black hole masses that exceed by 1-2 orders of magnitude the values obtained for these objects using techniques such as reverberation mapping and stellar velocity dispersion.  Moreover, the response of the double-peaked Balmer line profiles to fluctuations of the ionizing continuum and the shape of the Ly$\alpha$ profiles are incompatible with a SMBH binary. The binary broad-line region hypothesis is therefore disfavored. Other processes evidently shape these line profiles and cause the long-term velocity variations of the double peaks.
\end{abstract}

\keywords{
line: profiles ---
galaxies: active ---
galaxies: individual (1E 0450$-$1817, 3C 227, 3C 332, 3C 390.3, 3C 59,
Arp 102B, CBS 74, Mrk 668, Pictor A, PKS 0235+023, PKS 0921$-$213,
PKS 1020$-$103, PKS 1739+18) ---
quasars: emission lines
}

\section{Introduction}\label{S:intro}

Supermassive black hole (SMBH) binaries are thought to be a common, if not inevitable, outcome of the merger-driven evolution of galaxies \citep[e.g.]{Begelman1980,Volonteri2003,DiMatteo2005,Hopkins2006}. In the scenario described by \citet{Begelman1980}, after two galaxies merge, their central BHs sink into the merger core through dynamical friction on a time scale of $\sim 10^8$~years. The loosely bound binaries ($\sim$~kpc scale) later tighten to $\sim 1$ pc scale through the scattering of nuclear stars, until the ``loss cone''\footnote{Stars residing in this cone in phase space have the right combination of positions and momenta to interact with the binary and eventually take away the angular momentum through gravitational slingshots.} is depleted. Thereafter, gravitational wave driven angular momentum loss will not be significant until the separation is $10^{-2}$--$10^{-3}\;$pc. The difficulty in shrinking the orbit after it reaches $\sim$~1~pc is the so-called ``final parsec problem''. Without an efficient mechanism to remove angular momentum, SMBH binaries can stall at the pc scale for longer than the Hubble time. More recent calculations invoking interactions of the SMBH binary with a gaseous reservoir \citep[e.g.,][]{Armitage2002,Escala2004,Dotti2007,Hayasaki2007,Dotti2009b,Cuadra2009,Lodato2009} or more realistic stellar dynamical models \citep[such as non-spherical or rotating galaxies; e.g.,][]{Yu2002a,Merritt2004a,Khan2013} suggest that SMBH binaries can evolve towards a merger quickly because they lose angular momentum efficiently.

Finding sub-parsec scale SMBH binaries (or confirming their absence) is crucial in our understanding of galaxy evolution \citep[see recent reviews by][]{Popovic2012, Schnittman2013}. However, there is no direct evidence for existence of such close SMBH binaries, although a number of candidates have been reported.  Due to their close separation, we have little chance to spatially resolve the two BHs. There is only one close SBHB candidate that has been imaged with radio interferometers, CSO~0402$+$379, with a separation of 7~pc \citep{Maness2004,Rodriguez2006,Rodriguez2009}. All other current candidates have been suggested on the basis of indirect evidence.

Most of the proposed electromagnetic signatures of SMBH involve either radio jets or an accretion disk. For example, X-shaped radio jets seen in some galaxies are thought to reflect different spin directions of the two BHs \citep{Merritt2002, Zier2002}.  Periodic flares in the light curve, such as the ones seen in OJ 287 approximately every 12 years \citep{Sillanpaa1988, Lehto1996, Valtonen2006}, led the authors to propose a secondary BH plunging through the accretion disk of the primary BH. The nearby Seyfert galaxy NGC~4151 is a much more tentative case, suggested by \citet{Bon2012} on the basis of variability of its broad Balmer line profiles. Finally, thanks to the recent large sky surveys such as the Sloan Digital Sky Survey\footnote{\url{http://www.sdss.org/}} (SDSS), large systematic searches for spectroscopic signatures have become possible \citep[e.g.][]{Tsalmantza2011, Eracleous2012, Liu2013, Shen2013, Decarli2013}, namely, single or double-peaked emission lines, displaced from the host galaxy rest frame, and taken to indicate orbital motion in a SMBH binary \citep[see][]{Begelman1980,Gaskell1983,Gaskell1996}.

In this work we study the spectroscopic variability of 13 active galactic nuclei (AGNs) whose optical spectra display two displaced broad-line (BL) peaks, the so-called ``double-peaked emitters''\footnote{The line profiles do not always show two clearly separated peaks. In some cases the profiles are flat-topped or show two shoulders. Nonetheless, there is substantial historical precedent of using this term to describe these profile shapes collectively, therefore we adopt the term here as well.} (see examples in Figure~\ref{specs}).We concentrate on testing the SMBH binary interpretation for these line profiles. In our working scenario, each of the two BHs in the binary has an associated BL region, with the result that the orbital motion of the binary imparts a velocity separation between the lines from the two distinct BL regions \citep[see the heuristic model and illustrations in][]{Shen2010}. The separation of the peaks or shoulders in the observed line profiles suggests orbital separations of order 0.1~pc and orbital periods of order a few centuries, assuming BH masses of order $10^8~{\rm M}_\odot$ \citep[see, for example, Equations 1 and 2 in][]{Eracleous2012}.

In addition to the reasons mentioned above, testing the SMBH binary hypothesis for double-peaked emitters helps answer the question of the origin of these line profiles. Several scenarios have been suggested in the literature, including SMBH binaries and emission from the outer parts of the accretion disk around a single BH \citep[e.g.,][]{Chang1989,Chen1989,Eracleous1994,Strateva2003}. The accretion disk interpretation has passed many observational tests and does not suffer from many drawbacks, unlike the other candidate explanations \citep[a summary and critique of the scenarios can be found in][and Eracleous et al.\ 2009]{Eracleous2003}.\nocite{Eracleous2009} The analysis we present here provides the most comprehensive test of the SBHB hypothesis to date. The combination of this test and several other observations and physical arguments that we summarize in Section~\ref{discussion} render the binary black hole hypothesis an unlikely candidate for explaining double-peaked emission lines.
  
As long-term monitoring is critical for finding evidence of orbital motion in the variations of the double-peaked line profiles, we collect in this paper measurements of spectra of the these objects from the literature, dating back to the 1970s. We supplement the published data with measurements we make from newly acquired spectra and from spectra retrieved from the SDSS spectroscopic archive. Using all the available data, we search for changes in the radial velocities of the displaced peaks or shoulders that can be attributed to Keplerian motion of two BHs, following the method of \cite{Eracleous1997}. The collection of 13 objects we study here is substantially larger than the three objects studied by \cite{Eracleous1997}. Those three objects, Arp~102B, 3C~390.3, and 3C~332, are included in our present collection but their time series are now 2, 1.5, and 1.3 times longer, respectively. As a result, we obtain more stringent constraints on the orbital parameters of putative binaries in these objects.

In Section~\ref{obs}, we describe the published data as well as our new observations and data reduction. In Section~\ref{analysis}, we present the analysis of all available radial velocity measurements, and fit sinusoidal models to the velocity curves. We discuss the implications in Section~\ref{discussion}, where we also present additional arguments against the SMBH binary hypothesis based on a variety of other observations. We summarize our findings and present our conclusions in Section~\ref{summary}.

\section{Observations and Data Reduction}\label{obs}
We selected the 13 AGNs listed in Table~\ref{targetlist} and compiled data from \citet{Eracleous1997}, \cite{Gezari2007}, and \cite{Lewis2010}. The observations of the 13 objects by these authors span $\approx20$ years, from approximately 1985 to 2005. Their spectra were obtained using various telescopes of 1-m to 9-m aperture. We refer readers to these papers for detailed description of the observations and data reduction. We also retrieved additional historical spectra from the literature as well as from the SDSS. The sources of spectra for each object are listed in the last column of Table~\ref{targetlist}. 

\begin{table}
\caption[]{\label{targetlist} Target List}
\begin{tabular}{lllcc} 
\hline
Object	&	R.A. (J2000)			&	Decl. (J2000)			&	$z$     &      Refs.	\\
\hline													
 1E 0450$-$1817	&	04	52	35.91	&	$-$18	12	01.64	&	0.0590	&	1	\\
 3C 227	&	09	47	45.15	&	+07	25	20.60	&	0.0858	&	2,3,12,13	\\
 3C 332	&	16	17	42.52	&	+32	22	34.01	&	0.1517	&	2,12	\\
 3C 390.3	&	18	42	08.99	&	+79	46	17.13	&	0.0562	&	2--7,11	\\
 3C 59	&	02	07	02.18	&	+29	30	45.99	&	0.1100	&	1	\\
 Arp 102B	&	17	19	14.49	&	+48	58	49.43	&	0.0245	&	2,10,14	\\
 CBS 74	&	08	32	25.35	&	+37	07	36.26	&	0.0919	&	1,12	\\
 Mrk 668	&	14	07	00.39	&	+28	27	14.69	&	0.0766	&	2,10	\\
 Pictor A	&	05	19	49.72	&	$-$45	46	43.85	&	0.0340	&	1,8,9	\\
 PKS 0235+023	&	02	38	32.68	&	+02	33	49.67	&	0.2090	&	2	\\
 PKS 0921$-$213	&	09	23	38.89	&	$-$21	35	47.13	&	0.0534	&	1	\\
 PKS 1020$-$103	&	10	22	32.81	&	$-$10	37	44.37	&	0.1970	&	1	\\
PKS 1739+18	&	17	42	06.95	&	+18	27	21.06	&	0.1860	&	1	\\
\hline
\end{tabular}
{\it References.--}
(1) \cite{Lewis2010};
(2) \cite{Gezari2007};
(3) \cite{Osterbrock1976};
(4) \cite{Yee1981};
(5) \cite{Shafer1985};
(6) \cite{Popovic2012};
(7) \cite{Barr1980};
(8) \cite{Danziger1977};
(9) \cite{Carswell1984};
(10) \cite{Shapovalova2013};
(11) \cite{Gaskell1996};
(12) SDSS DR10;
(13) \cite{Netzer1982};
(14) \cite{Popovic2014}.
\medskip
\end{table}

For four of the targets, we carried out additional observations spanning the years 2009--2013. We obtained our spectra using the MDM Observatory\footnote{\url{http://mdm.kpno.noao.edu/}} Hiltner~2.4~m and McGraw-Hill~1.3~m telescopes. Spectra were taken with either the Boller and Chivens CCD Spectrograph (CCDS) or with the Modular Spectrograph (Modspec). CCDS observations were obtained with a 150~mm$^{-1}$ grating and a $1^{\prime\prime}$ slit, giving 7.6\,\AA\ resolution. Modspec observations used a 600~ mm$^{-1}$ grating and a $1^{\prime\prime}$ slit, with 3.4\,\AA\ resolution. One or multiple 900--1800~s exposures were taken for each target. The log of observations is given in Table~\ref{obslog}.

\begin{figure*}
\includegraphics[scale=0.45]{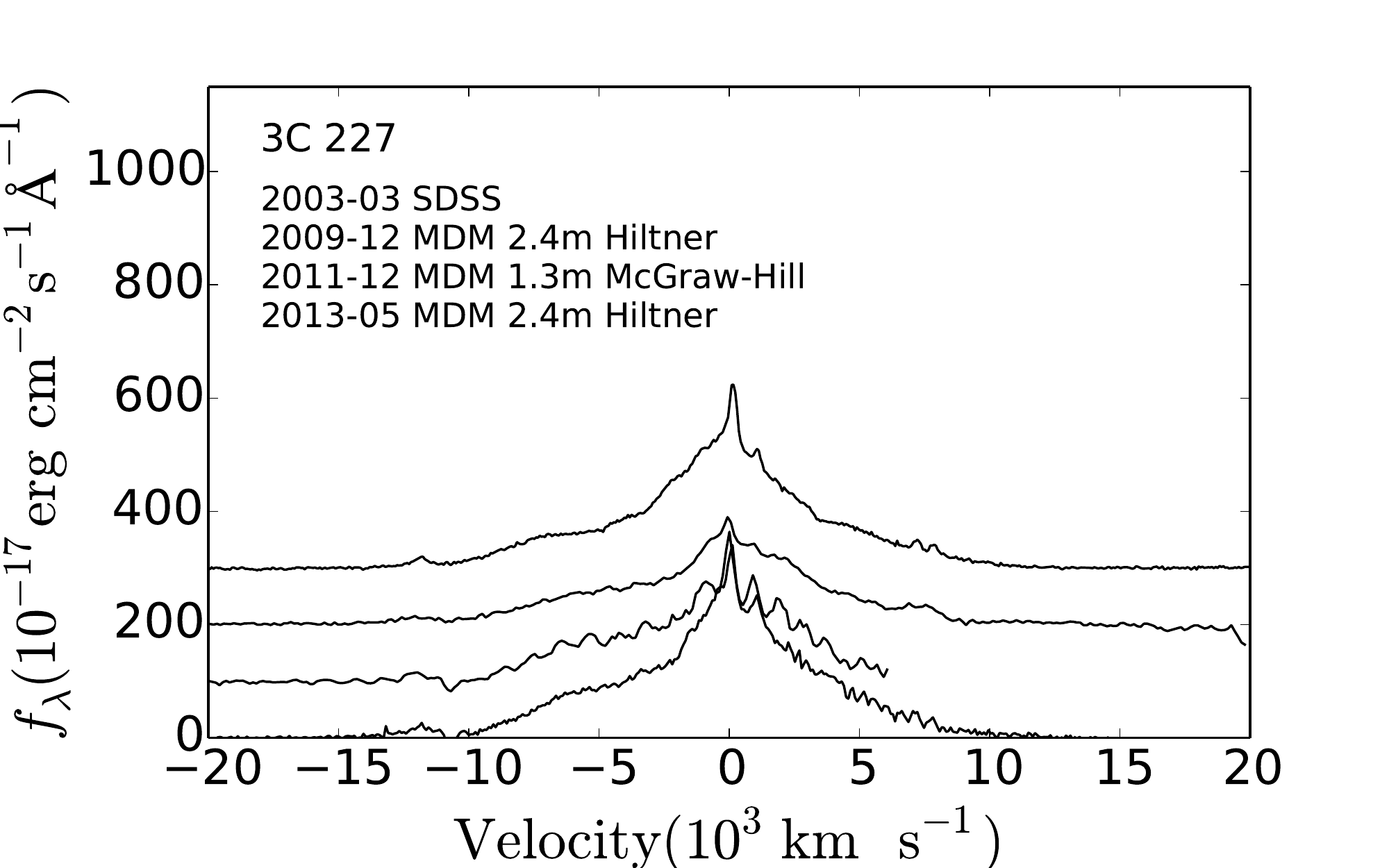}
\includegraphics[scale=0.45]{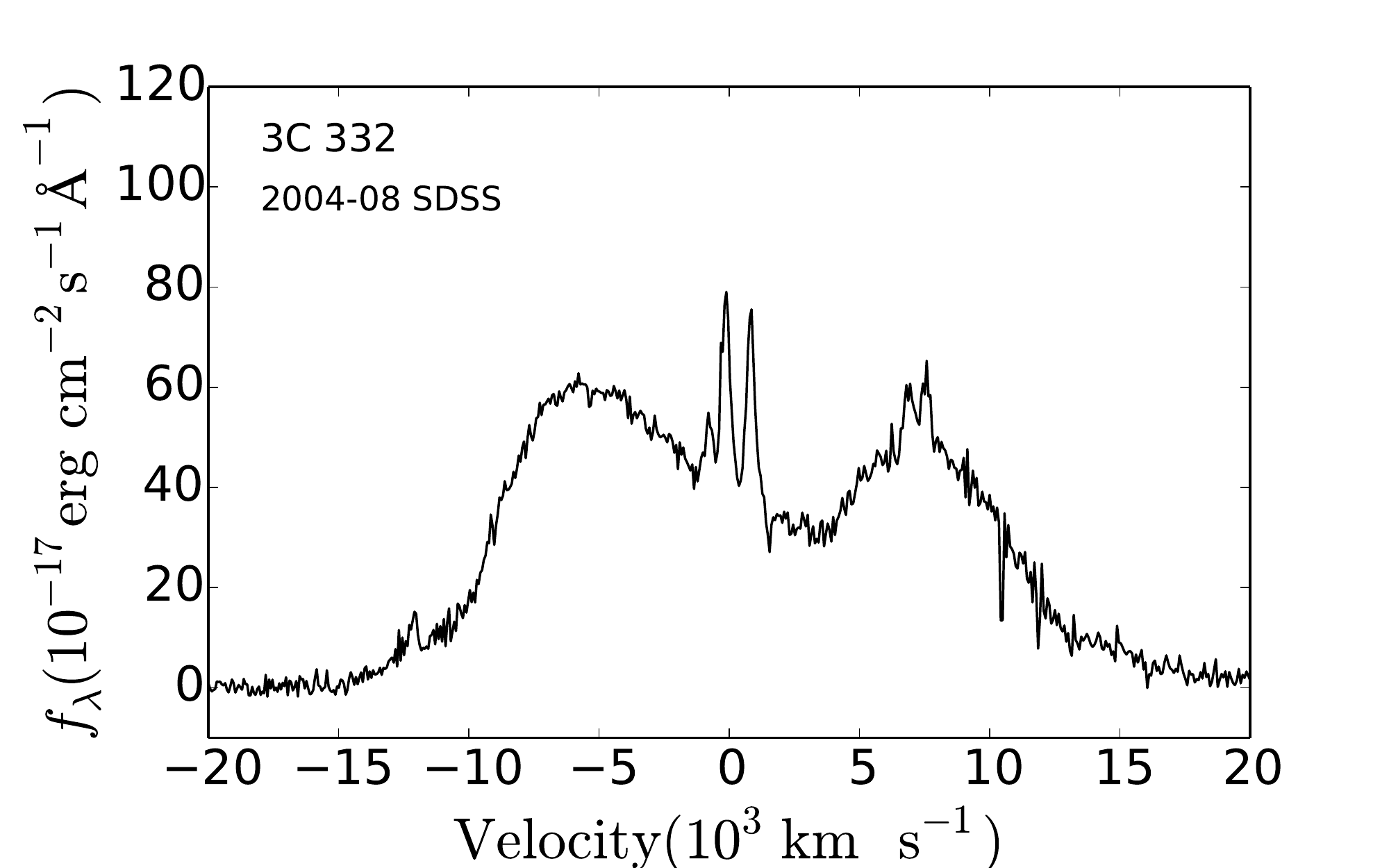}
\includegraphics[scale=0.45]{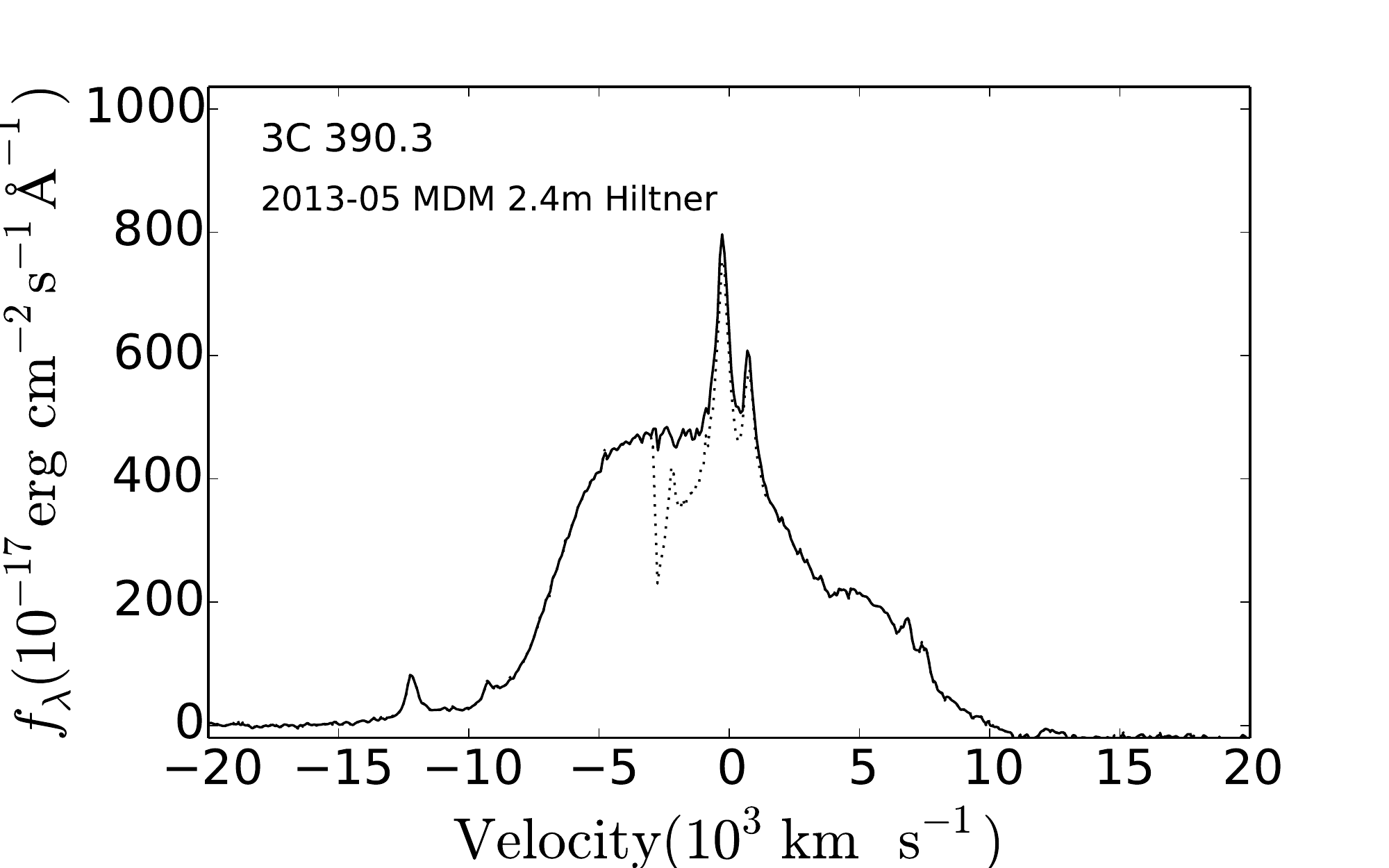}
\includegraphics[scale=0.45]{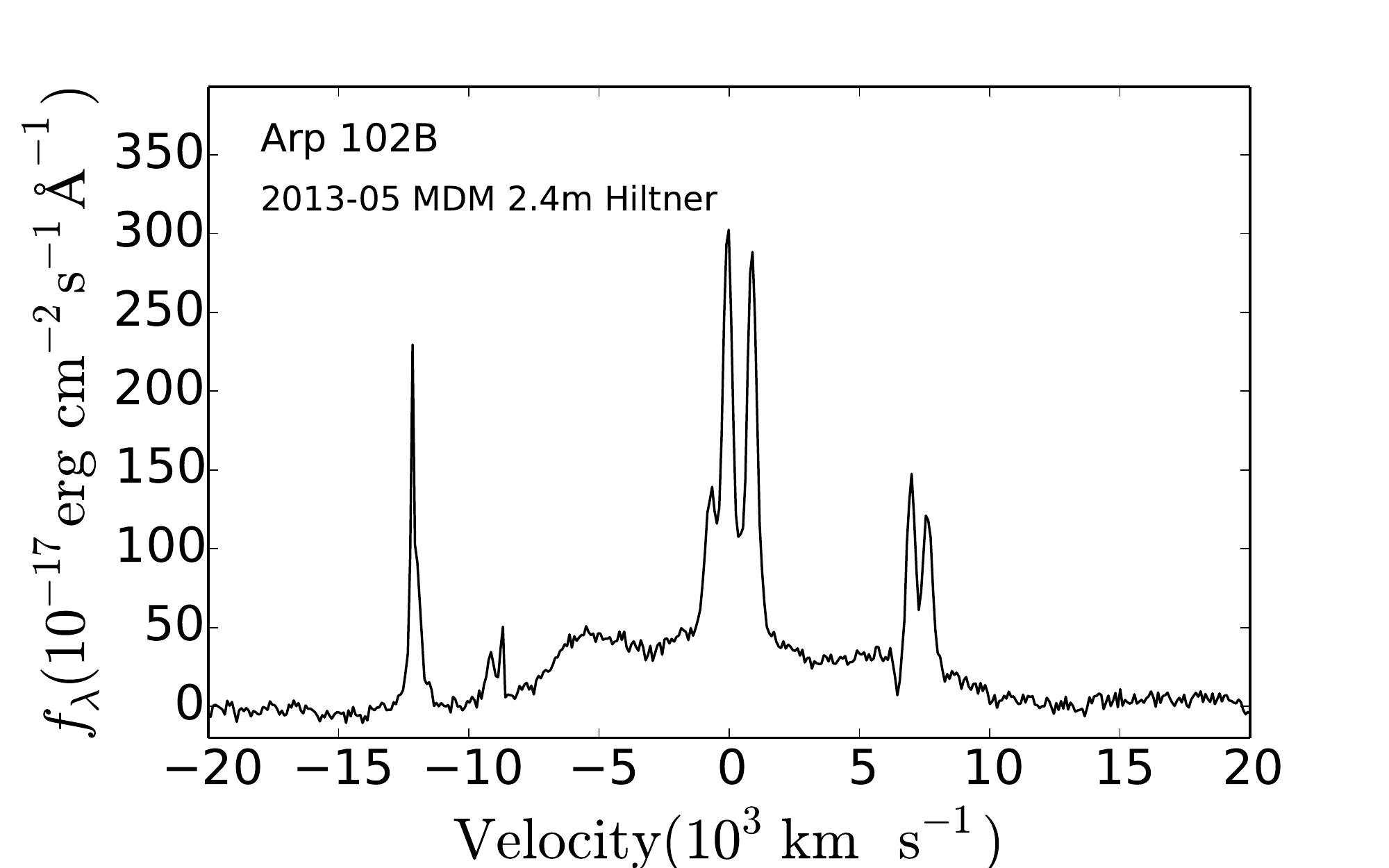}
\includegraphics[scale=0.45]{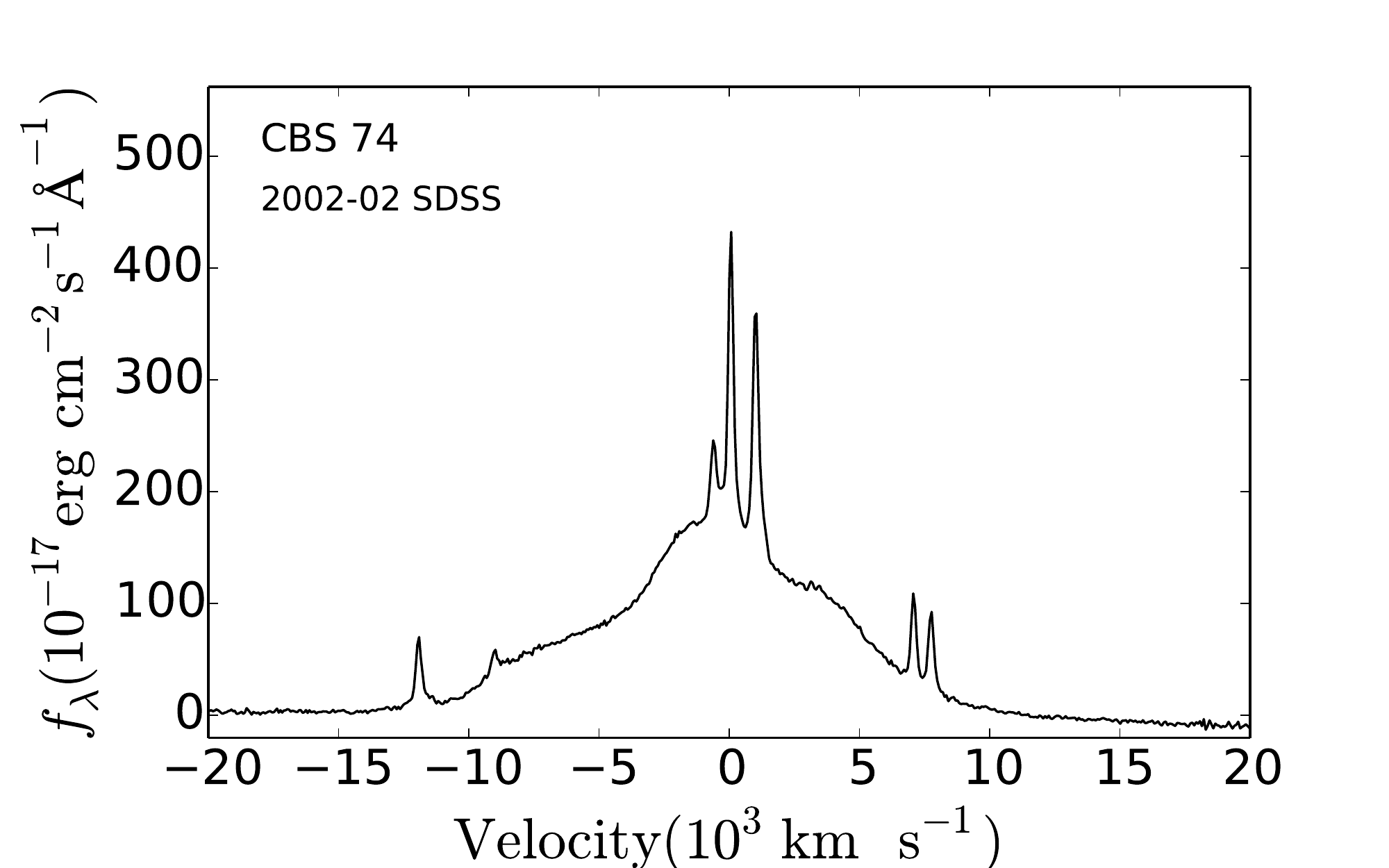}
\includegraphics[scale=0.45]{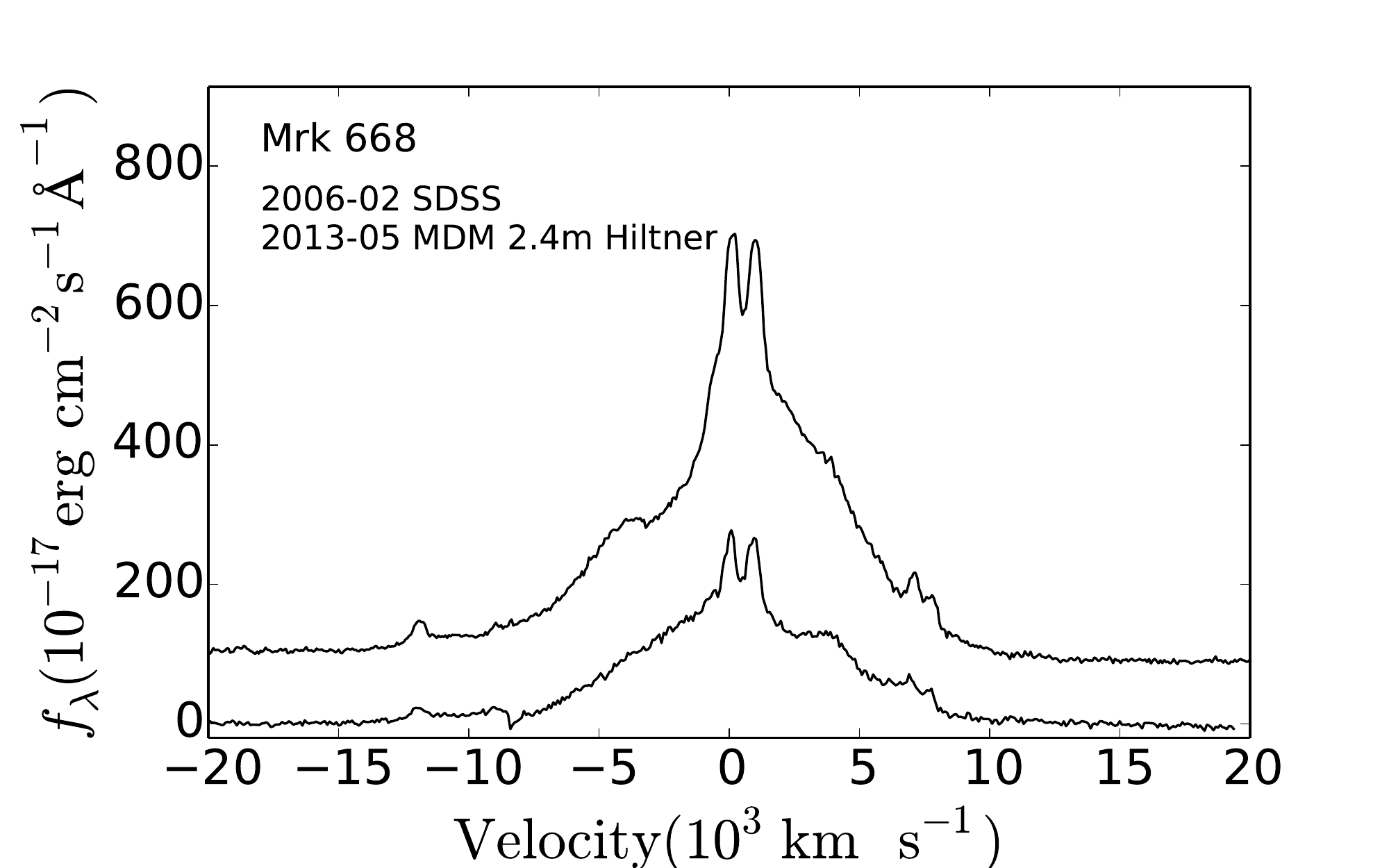}
\caption{\label{specs} Six H$\alpha$ spectra obtained during 2009--2013 at MDM Observatory, and four spectra from the SDSS. The dotted line in 3C~390.3 shows the original spectrum before atmospheric B-band correction. Vertical offsets have been applied for clarity.}
\end{figure*}

We reduced and calibrated the new spectra using the {\it Image Reduction and Analysis Facility}\footnote{\url{http://iraf.noao.edu/}} (IRAF).  For each object, the two-dimensional images were first bias and flat-field corrected. Then the one-dimensional spectra were extracted using the {\it apall} routine. Wavelength calibration was done using comparison lamps of Ar, Hg-Ne, and Xe with the {\it identify}, {\it reidentify}, and {\it dispcor} routines. The flux scale was calibrated using the spectra of standard stars by applying the {\it standard}, {\it sensfunc}, and {\it calibrate} routines. Finally, we correct the atmospheric B-band absorption for 3C~390.3 using the standard star Feige~34. Besides our follow-up observations, we also obtained four spectra from SDSS DR10 (see Table~\ref{targetlist}). Our reduced spectra and SDSS spectra are shown in Figure~\ref{specs}.

\section{Analysis}\label{analysis}

First, we follow the method described in Section~3.4 of \cite{Lewis2010} to measure the velocity of the shifted peaks of the H$\alpha$ BLs for all spectra for which such measurements have not already been made. We then fit sinusoidal curves to the red and blue velocity peaks simultaneously, and obtain the best-fit and the minimum orbital period for each object allowed by the secular curvature of the radial velocity curve.

\subsection{Line Profiles}\label{lines}

For each spectrum, we measure the velocities using the IRAF {\it splot} routine. At the region of each peak, we mark two continuum points and  fit a single Gaussian line profile. However, when the peak shapes are not Gaussian (for example, the flat-topped red peak of Mrk~668), the actual position of the peak can be quite uncertain. Hence, we perform multiple measurements using various continuum points and take the average to be the center.  We also take half of the difference between our measured minimum and maximum peak velocities, or 100 km s$^{-1}$ to be the error, whichever is larger. The H$\alpha$ narrow line (NL) is taken to be at zero velocity. The velocity of a shifted peak of the H$\alpha$ BL is calculated as $v=\left(\Delta\lambda/\lambda_{\rm NL}\right)c$, where $\Delta\lambda=\lambda_{\rm BL}-\lambda_{\rm NL}$ and $c$ is the speed of light.

\begin{table}
\caption{\label{obslog} MDM Observation Log}
\begin{tabular}{lllcc} 
\hline					
Object	&	Date (UT)	&	Instrument		&	Exposure (s)	\\
\hline								
3C 227	&	2009 Dec 16	&	2.4 m	CCDS	&	1200	\\
	&	2011 Nov 26	&	1.3 m	CCDS	&	$3\times1200$	\\
	&	2013 May 04	&	2.4 m	Modspec	&	1800	\\
3C 390.3	&	2013 May 03	&	2.4 m	Modspec	&	1800	\\
Arp 102B	&	2013 May 04	&	2.4 m	Modspec	&	900	\\
Mrk 668	&	2013 May 03	&	2.4 m	Modspec	&	1800	\\
\hline
\end{tabular}
\end{table}

For observations where we do not have access to the digital data (i.e., references~$3-11$ in Table~\ref{targetlist}), we convert the relevant published figure using {\it Plot Digitizer}\footnote{\url{http://plotdigitizer.sourceforge.net/}}, and estimate the velocity of the shifted peaks by eye. These estimates are certainly not as accurate as measurements done on numerical data, but give us valuable historical information on line shifts. However, only figures from \cite{Osterbrock1976}, \cite{Barr1980}, and \cite{Shapovalova2013}  have high enough resolution that we can measure the velocity confidently. \cite{Gaskell1996}, \cite{Gezari2007}, \cite{Lewis2010}, \cite{Popovic2012}, and \cite{Popovic2014} presented velocity shift measurements, which we use directly in our analysis. These authors used different methods to determine the peak location. \cite{Gaskell1996} used the Pogson method, which is very similar to fitting a Gaussian to the region of the line profile close to the peak; \cite{Gezari2007} measured the flux-weighted centroid of the top 10\% of the peak; \cite{Lewis2010} and \cite{Popovic2014} fitted a Gaussian profile to the peak; \cite{Popovic2012} used the parameters resulting from a disk model. The different measurements are fairly consistent when examined by eye, except for the noticeable difference between the redshifted velocity component in 3C390.3 measured by \cite{Gezari2007} and that by \cite{Popovic2012}. This is likely due to their different methods of removing the NLs. For consistency, we exclude the data of \cite{Popovic2012} from the periodicity test in Section~\ref{Ptest}.  We also note that \cite{Gaskell1996} measured the blue peak of H$\beta$ instead of H$\alpha$.

The measured velocities from the spectra obtained in this work and from the digitized figures are listed in Table~\ref{v_measured}. Radial velocity curves for each object are shown in Figure~\ref{vcurve}, and we examine them for the signature of orbital motion next. These measurements extend the previously published spectroscopic time series for Arp~102B, 3C~390.3, 3C~227, and Mrk~668 by factors of 1.7, 1.25, 2.7, and 1,6 respectively, which substantially influences the binary orbital parameters we derive below.

\begin{table}
\caption[]{\label{v_measured} Newly Measured velocities}
\begin{tabular}{llccc} 
\hline
Target   & Date (UT) & $U_1$ (blue)        & $U_2$ (red)         & Refs. \\
         &           & ($10^3$ km s$^{-1}$) & ($10^3$ km s$^{-1}$) & \\
\hline
3C 227	 & 1974 May/Dec	& $-2.04\pm0.10$   & ...           & 1  \\
	 & 2003 Mar	& ...		   & $0.95\pm0.10$ & 5	\\
	 & 2009 Dec 16	& $-1.5\pm0.10$	   & ...           & 2	\\
	 & 2011 Nov 26	& $-0.95\pm0.10$   & ...           & 2	\\
	 & 2013 May 04	& $-0.47\pm0.10$   & $0.96\pm0.10$ & 2	\\
3C 390.3 & 1974 May/Jun	& $-3.22\pm0.64$   & $4.67\pm0.69$ & 1	\\
	 & 1975 Jul	& $-4.00\pm0.50$   & ...           & 3	\\
	 & 2013 May 04	& $-4.18\pm0.29$   & $6.34\pm0.10$ & 2	\\
Arp 102B & 1998 Jul	& $-5.47\pm0.10$   & ...           & 4	\\
	 & 2003 Mar	& $-5.47\pm0.10$   & ...           & 4	\\
	 & 2006 Aug	& $-5.56\pm0.10$   & ...           & 4	\\
	 & 2013 May 04	& $-5.31\pm0.10$   & $6.74\pm0.10$ & 2	\\
Mrk 668	 & 2006 Feb 27	& $-4.48\pm0.10$   & $3.79\pm0.10$ & 5	\\
	 & 2013 May 03	& ...              & $3.83\pm0.10$ & 2	\\
CBS 74	 & 2002 Feb 06	& $-2.15\pm0.10$   & $3.86\pm0.10$ & 5	\\
3C 332	 & 2004 Aug 22	& $-5.88\pm0.10$   & $7.90\pm0.10$ & 5  \\
\hline
\end{tabular}
{\it References.--}
(1)	\cite{Osterbrock1976};
(2)	This work;
(3)	\cite{Barr1980};
(4)	\cite{Shapovalova2013};
(5)	SDSS.
\end{table}

\subsection {Tests for Orbital Motion}\label{Ptest}

To derive constraints on the total mass of a hypothesized SMBH binary, we follow the method of \citet{Eracleous1997}. In summary, we adopt a circular orbit model and fit the observed radial velocities $u_1$ and $u_2$ of the two peaks with the following model
\begin{eqnarray} \label{u12}\nonumber
u_1(t)=-(v_1\; \sin i)\; \sin \left[{2\pi\over P} (t-t_0) \right] ~{\rm and}\\
u_2(t)=(v_2\; \sin i)\; \sin \left[{2\pi\over P} (t-t_0)\right]\;, \qquad
\end{eqnarray}
where $v_1, v_2$ are the true (i.e. tangential) orbital velocities, $i$ is the inclination angle, $P$ is the period, and $t_0$ is the time at which the two peaks coincide at zero radial velocity (conjunction). The assumption of circular orbits is justified by the results of stellar dynamics simulations that follow the evolution of the binary following the merger of the two parent galaxies. As a rule, for the eccentricity to grow appreciably it has to start out with values, $e\gs 0.5$, and the binary mass ratio should also be large, $q=M_1/M_2 \gs 10$ \citep[e.g.,][]{Quinlan1996,Sesana2006,Wang2014,Vasiliev2015}. As we note below, the velocity ratio of the two peaks in the double-peaked profiles of our sample imply small mass ratios, $q=1$--2, which would suppress the growth of eccentricity. Most relevant to the problem we are concerned with here are the N-body simulations of merging galaxies by \citet{Khan2013} and \citet{Holley2015}, which find that the eccentricity of the binary quickly decays to zero for galaxies with no rotation or net prograde rotation. For galaxies with retrograde rotation, the eccentricity is quickly driven to unity and the binary merges as soon as it hardens.

To find the best-fitting model parameters, we use two schemes. In the first scheme, we minimize the $\chi^2$ statistic, given by
\begin{equation}
\chi^2=\sum_{i=1}^{N_1}{\left(U_1^i-u_1^i\over \sigma_1^i\right)^2}+\sum_{i=1}^{N_2}{\left(U_2^i-u_2^i\over\sigma_2^i\right)^2}
\label{eq:chisq}
\end{equation}
where $U_1, U_2$ are the observed velocities, $u_1, u_2$ are models from equation~\ref{u12}, and $N_1,N_2$ are the number of data points for the two velocity peaks. Note that uncertainties in velocity mainly come from line profile changes, possibly due to either reverberation or dynamical effects \citep[e.g.,][respectively]{Barth2015,Eracleous1997}, rather than measurement errors such as those in Table~\ref{v_measured}. Therefore, we set the error bar, $\sigma$, to be the standard deviation of multiple measurements within a year. When there is only one observation available in a year, we use the average error from other epochs where multiple measurements of the same object are available. In practice, we scan through the three-dimensional parameter space $(t_0,\; v_1 \sin i,\; v_2 \sin i)$  for $P\in [5, 10^4]$ years with 100 logarithmically spaced values of $P$. The best-fitting model parameters obtained by this scheme are summarized in Table~\ref{P_fitting_chi}, where we also include the reduced $\chi^2$ values for the best fits, $\chi^2_{\rm\nu,min}$, and the number of degrees of freedom. In several cases, we were not able to find the minimum $\chi^2$ in the range of periods that we tested because $\chi^2$ decreases monotonically up to the maximum period we consider. Therefore, we conclude that $P > 10^4\;$yr and we indicate this conclusion in Table~\ref{P_fitting_chi}. As one can see in Table~\ref{P_fitting_chi}, the values $\chi^2_{\rm\nu,min}$ are significantly larger than unity, indicating that the models do not provide a satisfactory description of the data. In the case of PKS~0235+023, the probability that the binary orbit model is a suitable model is 0.06 while in all other cases this probability is less than $8\times 10^{-5}$. Even though a better fit may be possible for periods greater than $10^4\;$yr for some objects, such periods lead to unreasonably high black hole masses, as we show later in this section.

In the second scheme we minimize the {\it unweighted} squared deviations of the model from the data. In other words, we minimize the statistic of equation~(\ref{eq:chisq}) after setting $\sigma_{1,2}^i$ to unity for all data points. This scheme amounts to accepting that there will be some scatter in the individual measurements as a result of profile variability on times scales of order a few years. In Figure~\ref{P_fitting_lsq} we show the best-fitting models from this scheme as solid lines superposed on the data. The best-fitting model parameters are listed in Table~\ref{P_fitting_lsq}. The periods obtained by the unweighted least squares method are similar to those obtained by the $\chi^2$ method.  We also note that the periods allowed for each object are much longer than the span of our monitoring program, rendering any ``periodicity'' found in this analysis unreliable. In column~7 of Table~\ref{P_fitting_lsq}, we give the expected year of the next conjunction, according to the best-fitting model. At the time of conjunction, according to the binary model, the two peaks will overlap and the emission line profile will appear single-peaked and have its minimum width.

\greensout{and in Table 5 for minimum allowed period. The minimum period is drawn from the 99\% confidence contour, having $\chi^2=\chi_{\rm min}^2+\Delta\chi^2$ with $\Delta\chi^2=11.3$ or 9.21 for $N_{\rm par}=3$ or $N_{\rm par}=2$, respectively \hbox{\citep{Lampton1976}}. The reduced chi-square listed in the Tables is $\chi_{\nu}^2=\chi^2/(N_1+N_2 - N_{\rm par})$, where $N_1,N_2$ are the number of data points for the two velocity peaks, $N_{\rm par}=3$ for objects with both red and blue peaks measured, and $N_{\rm par}=2$ for those with only one peak visible.  These are conservative choices of  $\Delta\chi^2$ in the sense that $t_0$ is not a physically ``interesting'' parameter. On the other hand, since $\chi_{\nu, \rm min}$ is significantly greater than 1 in all cases, the binary model is not a good description of the data.  Either we have significantly underestimated the error bars in velocity, or the orbiting binary black hole hypothesis is not valid. We believe that our empirical definition of the error bars based on the scatter in measured radial velocities over the multiple measurements in the same year does capture the uncertainty in the locations of the peaks. Therefore, we favor the latter explanation for the large values of $\chi_{\nu, \rm min}$. To reinforce our point, we note that in 3C~332 and 3C~390.3, the peaks of the H$\alpha$ line are fairly easy to discern and measure (see Figure~\ref{specs}), yet the radial velocity curves show very significant deviations from the orbital model (See Figure~\ref{vcurve}).} 

We also searched for periodic signals in both velocities and fluxes using the Lomb-Scargle periodogram \citep{Lomb1976, Scargle1982}\footnote{For velocities, we use all the observations available except for the data points from \cite{Popovic2012}. For fluxes, we only use the measurements by \cite{Gezari2007} and \cite{Lewis2010}.}. The Lomb-Scargle analysis is similar to the ordinary power spectrum, but particularly suitable for unevenly distributed data points. It also has the advantage of finding the phase automatically. Similar to what we found using sinusoidal fits, the periodogram did not find any significant period for our objects that is within the length of monitoring span.

We conclude that for the objects of interest in this work, either the periods are much longer than 20 years, or mechanisms other than orbiting BHs are responsible for the emission-line velocity changes.

Notwithstanding the significant deviations of the observed radial velocity curves from the orbital models, we go on to explore the consequences of the \bluesout{minimum} orbital periods implied by the fits. Assuming the two BHs have masses $M_1$ and $M_2$, with a mass ratio $q=M_1/M_2 \geq 1$, we obtain a lower limit on the total mass of the binary $M=M_1+M_2$ using the following relations
\begin{equation}
M>4.7 \times 10^8 \left(1+q\right)^3 \left({P\over {\rm 100 \;yr}}\right)
\left({v_1 \sin i\over 5000 {\rm \;km\;s^{-1}}} \right)^3 \Msol
\end{equation}
or
\begin{equation}
M>4.7 \times 10^8 \left({1+q\over q}\right)^3
\left({P\over {\rm 100 \;yr}}\right)  \left({v_2 \sin i \over 5000 {\rm
\;km \;s^{-1}}} \right)^3\Msol.
\end{equation}
\citep[see][]{Eracleous1997}. The mass ratio is inferred from the best-fitting velocity amplitudes through $q=v_2/v_1$. For objects with only one peak visible, even though it is not possible to estimate $q$ we can still put a constraint on the total mass since $q\geq1$ by definition; therefore, $(1+q)^3\geq8$ and $[(1+q)/q]^3\geq1$. The values for total mass of the SMBH binary that we obtain by the this approach are listed in Tables~\ref{P_fitting_chi} and \ref{P_fitting_lsq}.

\begin{table*}[t]
\begin{center}
\caption[]{\label{P_fitting_chi} Best-fit binary orbit parameters from the $\chi^2$ method}
\begin{tabular}{lcccccc} 
\hline													
	&	$P$	&	$v_1 \sin i$	&	$v_2 \sin i$	&		&		&		\\
Object	&	(yr)	&	($10^3$ km s$^{-1}$)	&	($10^3$ km s$^{-1}$)	&	$q$	&	$\log(M/M_\sun)$	&	$\chi_{\nu,\rm min}^2$ (d.o.f.)	\\
(1) &	(2)	&	(3)	&	(4)	&	(5)	&	(6)	&(7)	\\
\hline													
1E 0450.3--1817 &	37	&	3.34	&	5.56	&	1.66	&	9.0	&	3.41 ~(33)	\\
3C 227	        &	171	&	...	&	2.29	&	...	&     $>7.9$	&	6.42 ~(23)	\\
3C 332	        &	158	&	8.4	&	10.66	&	1.27	&      10.6	&	5.77 ~(59)	\\
3C 390.3	&	$>10^4$	&	30.63	&	37.10	&	1.21	&    $>14.1$	&	7.66 ~(99)	\\
3C 59	        &	79	&	2.11	&	4.11	&	1.95	&	8.9	&	6.14 ~(21)	\\
Arp 102B	&	$>10^4$	&	7.96	&	8.55	&	1.07	&    $>12.2$	&	3.09 (176)	\\
CBS 74	        &	79	&	1.9	&	4.00	&	2.11	&	8.8	&	2.77 ~(29)	\\
Mrk 668	        &	$>10^4$	&	72.57	&	...	&	...	&    $>14.2$	&	9.92 ~(21)	\\
Pictor A	&	136	&	5.1	&	5.49	&	1.08	&	9.8	&	4.11 ~(17)	\\
PKS 0235+023	&	232	&	...	&	5.56	&	...	&     $>9.2$	&	1.63 ~(14)	\\
PKS 0921--213	&	$>10^4$	&	16.59	&	16.64	&	1.00	&    $>14.1$	&	2.47 ~(24)	\\
PKS 1020--103	&	$>10^4$	&	24.39	&	32.45	&	1.33	&    $>13.8$	&	4.94 ~(15)	\\
PKS 1739+18	&	79	&	2.79	&	3.60	&	1.29	&	8.9	&	4.15 ~(31)	\\
\hline
\end{tabular}
\end{center}
\vskip -6pt
\centerline{\vbox{\hsize 5truein
{\it Table Columns.--} (1) object name, (2) best-fitting orbital period, (3) projected velocity amplitude of primary, (4) projected velocity amplitude of secondary, (5) mass ratio ($q=v_2/v_1$), (6) total mass of binary implied by best-fitting model, (7) reduced $\chi^2$ corresponding to best fit and number of degrees of freedom.
}}

\begin{center}
\caption[]{\label{P_fitting_lsq} Best-fit binary orbit parameters from the unweghted least squares method}
\begin{tabular}{lcccccc} 
\hline
Object	&	$P$	&	$v_1 \sin i$	&	$v_2 \sin i$	&		&		&	\\
&	(yr)	&	($10^3$ km s$^{-1}$)	&	($10^3$ km s$^{-1}$)	&	$q$	&	$\log(M/M_\sun)$	&$t_0$	\\
(1) &	(2)	&	(3)	&	(4)	&	(5)	&	(6)	&(7)	\\
\hline													
1E 0450.3--1817	&	54	&	3.29	&	4.82	&	1.47	&	9.0	&	2037	\\
3C 227	        &	158	&	...	&	2.26	&	...	&     $>7.8$	&	2097	\\
3C 332	        &	136	&	7.56	&	10.23	&	1.35	&      10.5	&	2089	\\
3C 390.3	&	$>10^4$	&	33.45	&	39.03	&	1.17	&    $>14.2$	&	...	\\        
3C 59	        &	79	&	2.37	&	4.16	&	1.76	&       8.9	&	2060	\\
Arp 102B	&	$>10^4$	&	5.65	&	5.75	&	1.02	&    $>11.7$	&	...	\\        
CBS 74	        &	100	&	1.75	&	3.75	&	2.14	&	8.8	&	2069	\\
Mrk 668	        &	$>10^4$	&	18.58	&	...	&	...	&    $>12.4$	&	...	\\        
Pictor A	&	$>10^4$	&	8.96	&	9.35	&	1.04	&    $>12.4$	&	...	\\        
PKS 0235+023	&	171	&	...	&	4.85	&	...	&     $>8.9$	&	2139	\\
PKS 0921--213	&	$>10^4$	&	6.99	&	7.01	&	1.00	&    $>12.0$	&	...	\\        
PKS 1020--103	&	$>10^4$	&	11.71	&	15.6	&	1.33	&    $>12.9$	&	...	\\        
PKS 1739+18	&	100	&	2.66	&	3.48	&	1.31	&	8.9	&	2070	\\
\hline
\end{tabular}
\end{center}
\vskip -6pt
\centerline{\vbox{\hsize 5truein
{\it Table Columns.--} (1) object name, (2) best-fitting orbital period, (3) projected velocity amplitude of primary, (4) projected velocity amplitude of secondary, (5) mass ratio ($q=v_2/v_1$), (6) total mass of binary implied by  best-fitting model, (7) expected year of next conjunction (i.e., time of vanishing projected velocities).
}}
\end{table*}

\section{Discussion of Results}\label{discussion}
\renewcommand{\thefigure}{\arabic{figure}}

\begin{figure*}
\begin{center}
\begin{flushleft}
\includegraphics[scale=0.43]{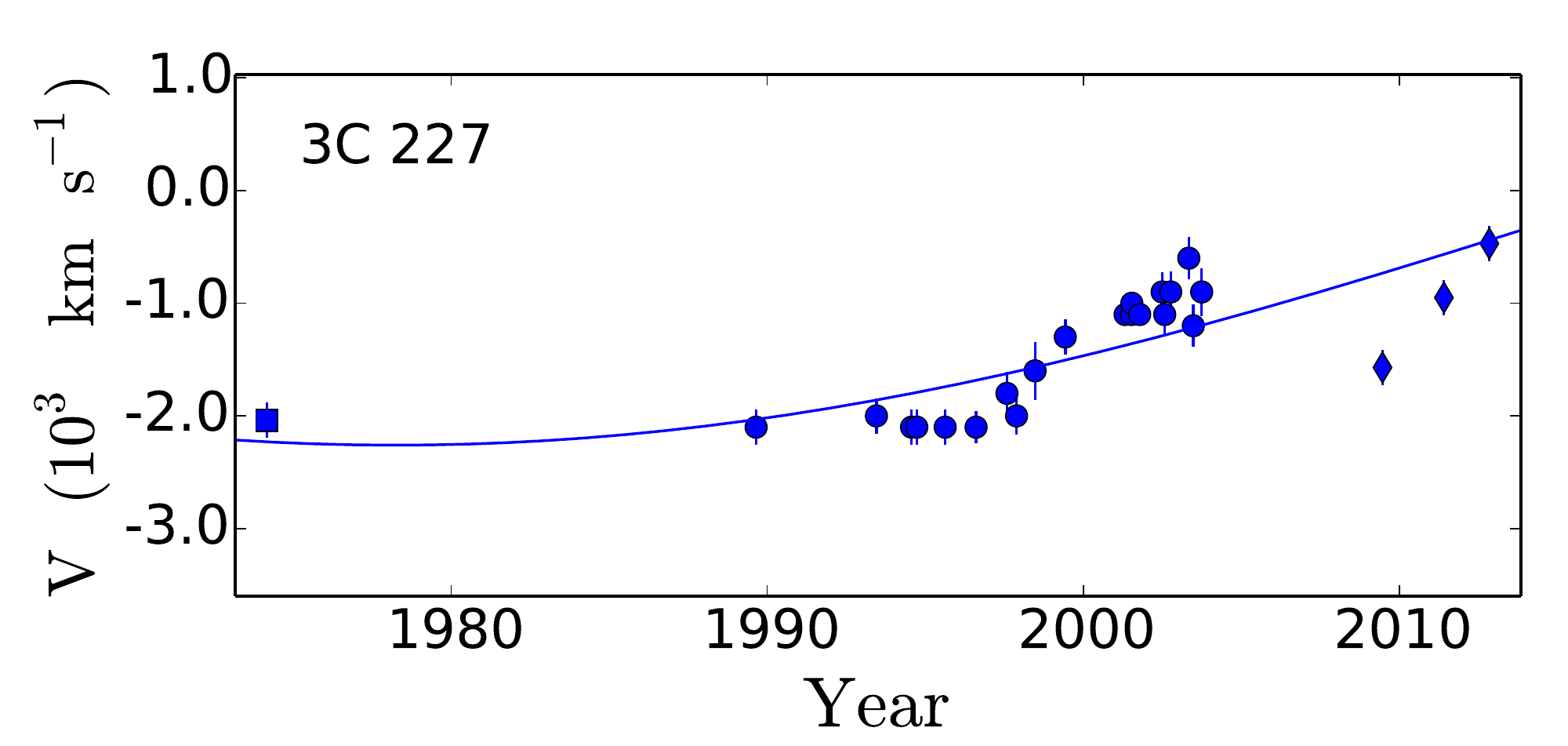}
\includegraphics[scale=0.43]{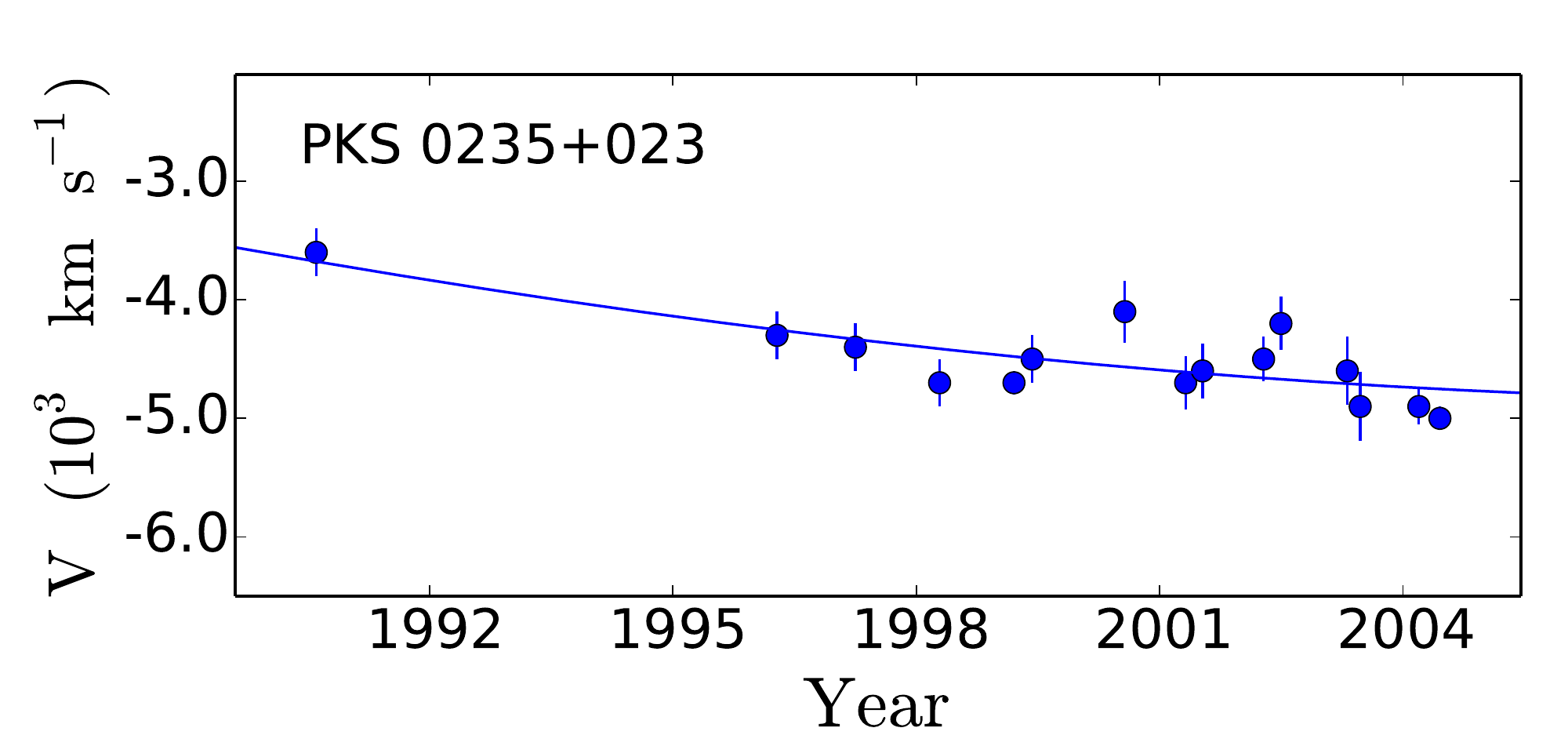}
\includegraphics[scale=0.43]{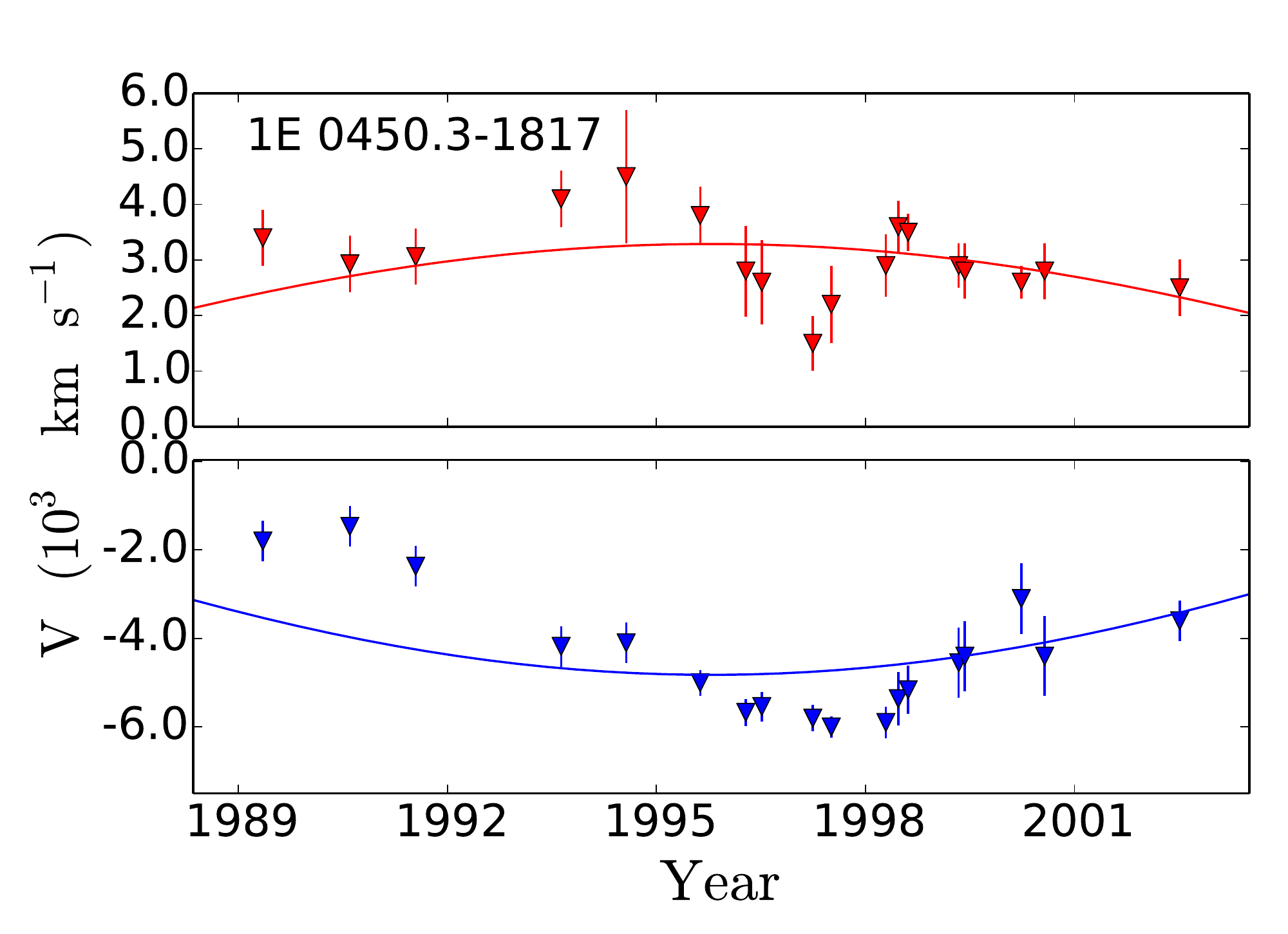}
\includegraphics[scale=0.43]{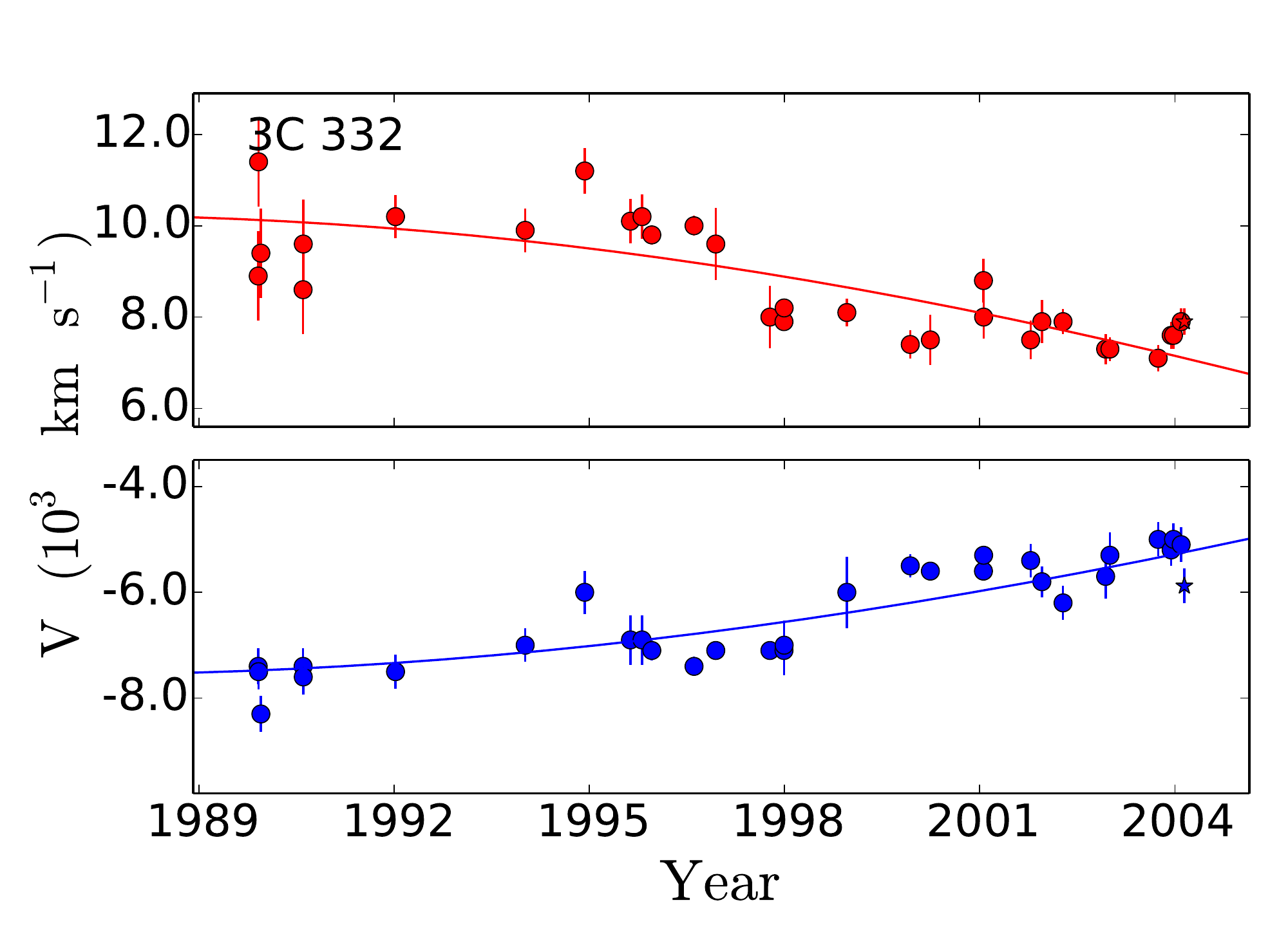}
\includegraphics[scale=0.43]{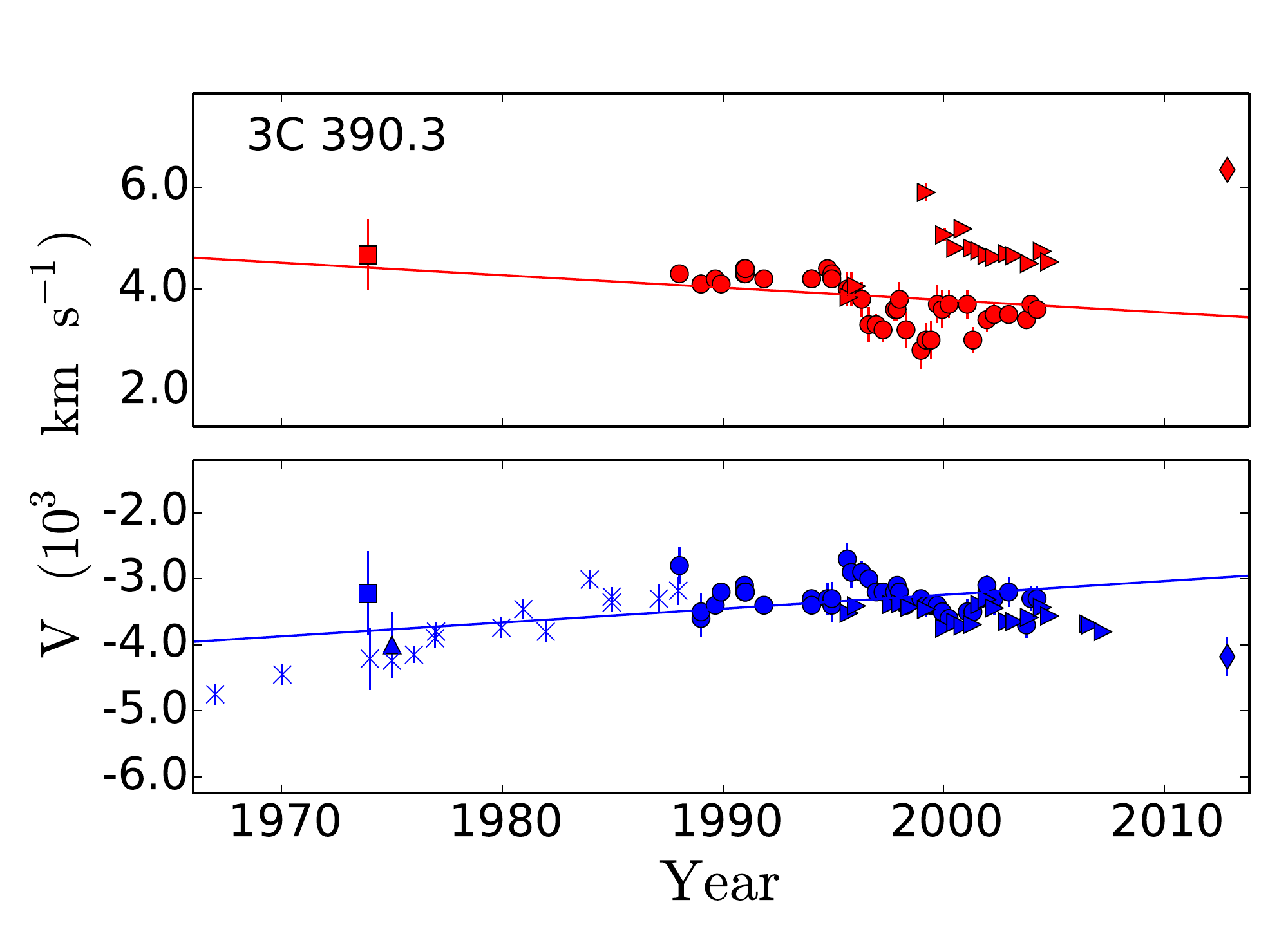}
\includegraphics[scale=0.43]{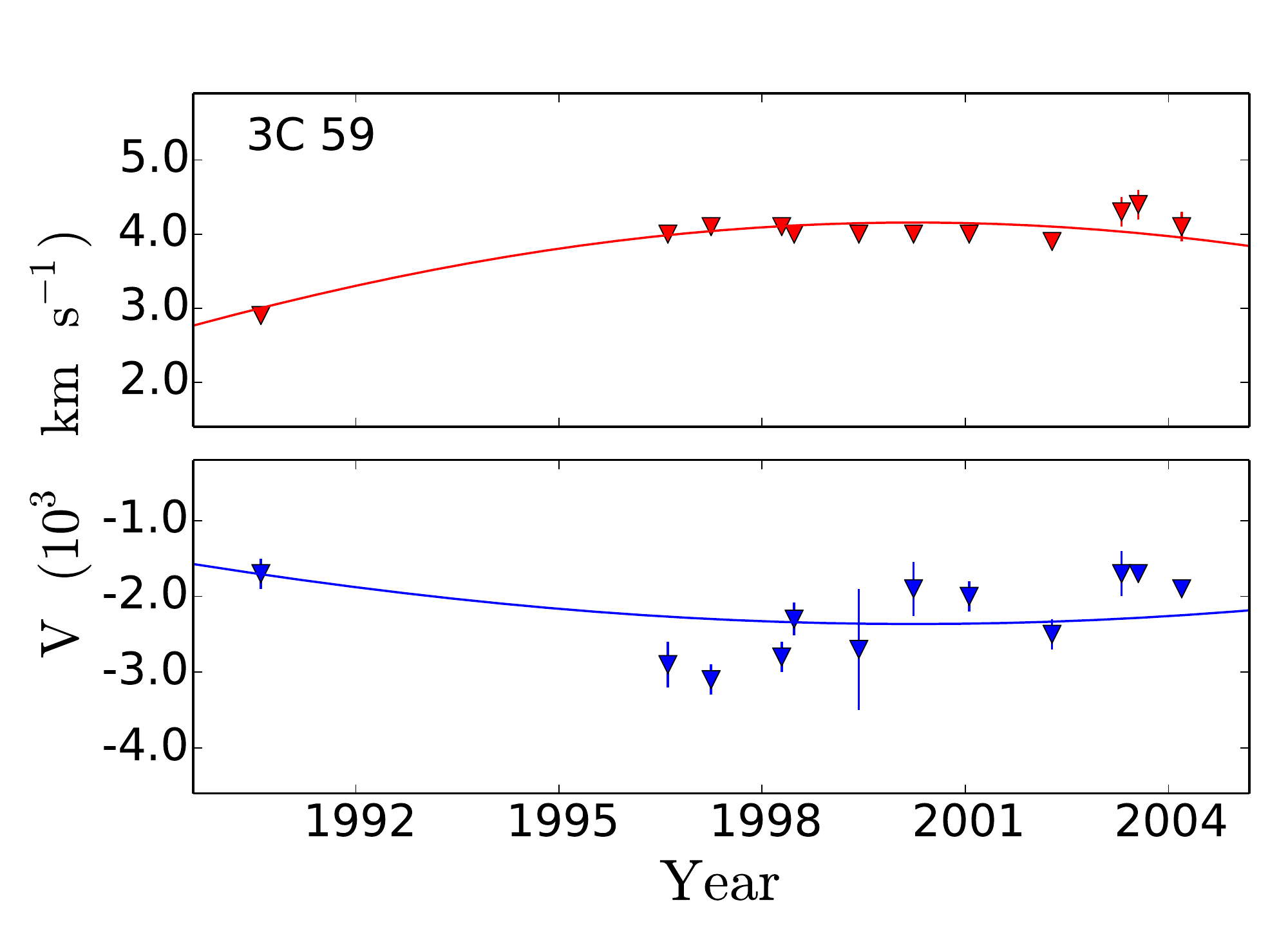}
\includegraphics[scale=0.43]{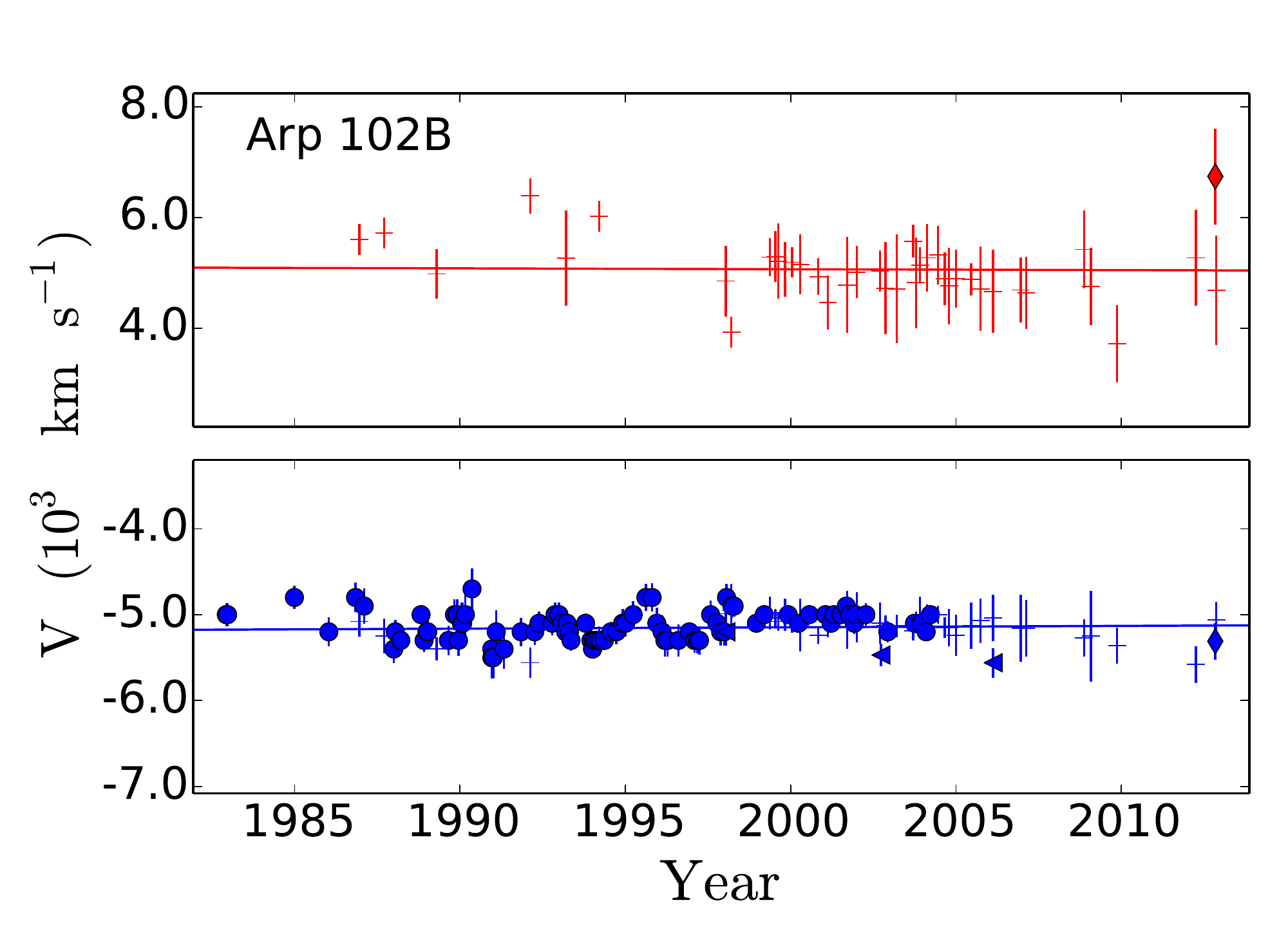}
\includegraphics[scale=0.43]{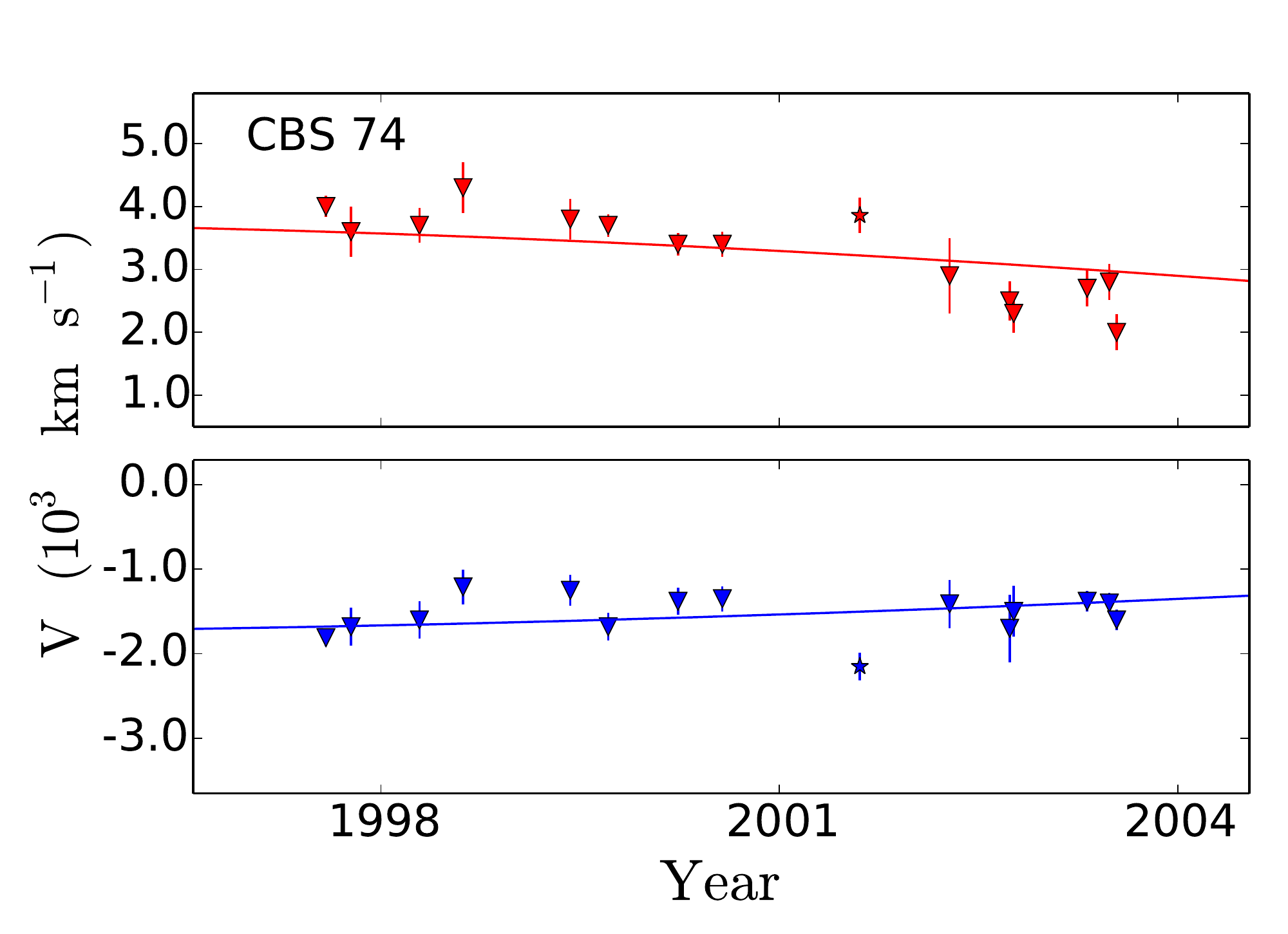}
\caption{\label{vcurve}
Velocities of the shifted peaks of the broad H$\alpha$ line and fitted curves. 
Circle: \cite{Gezari2007};
Down-triangle: \cite{Lewis2010};
Square: \cite{Osterbrock1976};
Up-triangle: \cite{Barr1980};
Diamond: This work;
``x'': \cite{Gaskell1996};
Left-triangle: \cite{Shapovalova2013};
Right-triangle: \cite{Popovic2012};
Plus: \cite{Popovic2014};
Star: SDSS.
Solid curves are the best-fit models from the unweighted least squares method (Table~\ref{P_fitting_lsq}).
\bluetext{Note that the velocity scale on the vertical axis has been expanded to show detail (i.e., it does not start at zero).}
}
\end{flushleft}
\end{center}
\end{figure*}

There have been many similar attempts to find SMBH binaries using displaced broad peaks, but none have succeeded. \cite{Gaskell1996} showed velocity shifts in the spectrum of 3C~390.3 that could be fitted with a 300 year period. However, \cite{Eracleous1997} later rejected this interpretation, as they found that the motion in 3C~390.3 stopped after 1988 (which is also evident from our Figure~\ref{vcurve}).  Lacking clear spectroscopic evidence for orbital motion in the velocity curves, the three objects chosen for study by \cite{Eracleous1997}, 3C~390.3, Arp~102B, and 3C~332, were found to require masses $>10^{10}\,M_{\odot}$ under the binary BL region hypothesis, and our extension of that monitoring has only increased those limits.  Such masses are far in excess of those measured using direct methods such as reverberation mapping, from which \cite{Shapovalova2013} estimated Arp~102B to have $M=1.1\times10^8\Msol$, and \cite{Sergeev2011} estimated 3C~390.3 to have $M=2.0\times10^9\Msol$, both much less than $10^{12}\Msol$ and $10^{14}\Msol$, respectively, obtained under the binary BL region assumption (see Tables~\ref{P_fitting_chi} and \ref{P_fitting_lsq}).

BH mass estimates for several of the objects in this study were also made by \citet{Lewis2006} using the velocity dispersion in the \ion{Ca}{2} infrared triplet and the correlation between BH mass and stellar velocity dispersion.  The results range from $4\times10^7\,M_{\odot}$ for 1E~0450.3$-$1817, Pictor~A, and PKS~0921$-$213, to $1.1\times10^8\,M_{\odot}$ for Arp~102B, and $5\times10^8\,M_{\odot}$ for 3C~390.3.  These are all a few orders of magnitude smaller than the masses derived under the binary BL region assumption (listed in Tables~\ref{P_fitting_chi} and \ref{P_fitting_lsq}), a further indication that the latter hypothesis is \bluesout{unreasonable} untenable.

We offer the following additional arguments against the SMBH binary interpretation of double-peaked emission lines.

\begin{enumerate}

\item
The profiles of the Ly$\alpha$ lines of double-peaked emitters, are not double-peaked. Instead they have a single peak that is located close to the systemic redshift, i.e., between the two peaks of the broad Balmer lines \citep[]{Halpern1996,Eracleous2009}. These argue strongly against any scenario that attributes the peaks of the Balmer lines to separate and physically distinct gaseous regions that move relative to each other.

\item
In the case of 3C~390.3, we can use constraints on the inclination angle of the radio jet, assumed to be perpendicular to the orbital plane, to refine the constraint on the mass of a hypothesized SMBH binary: \citet{Eracleous1996} estimated that $i > 19^\circ$ based on the observed superluminal motion in the radio jet. Applying this constraint raises the minimum mass of the binary to $\log(M_{\rm min}/M_\odot)=14.6$, which is rather implausible. Moreover, the binary separation in 3C~390.3 implied by the constraints is 28~pc. This is in \bluesout{sharp} contradiction with the results of reverberation mapping of 3C~390.3 that show the two peaks to respond to variations of the continuum virtually simultaneously \citep[within 3 days of each other, much shorter than the light crossing time corresponding to $>28$~pc; see][]{Dietrich2012}. Similarly, in the case of Arp~102B, the binary separation implied by the available constraints is $>6$~pc. In contrast the entire broad H$\alpha$ line responds to continuum changes within 22~days \citep{Shapovalova2013} with a negligible lag between the blue and red sides \citep{Popovic2014}, i.e., far from what one would expect for two widely-separated BL regions.

\item
  The working hypothesis on which the radial velocity test for a SMBH binary is based is that the two members of the binary are moving relative to the rest frame of the host galaxy and carrying with them the gas in the BL region \citep[see for example, the models of][for a possible geometry]{Hayasaki2007,Cuadra2009}. The width of the emission lines from the binary is dominated by the individual BL region rather than by the orbital velocities of the BH binary. The projected velocity separation between the two BLs should be considerably smaller than their widths based simply on Kepler's laws. This was pointed out by \citet[][see their footnote 3]{Chang1989} and more recently demonstrated by \citet{Shen2010} using a heuristic model for the BL region. Thus, we should not expect the combined line profile from a SMBH binary to show two well separated peaks as is the case for several of the objects in our sample (e.g., Arp~102B, 3C~332, Pictor~A, PKS~0921--21).

\end{enumerate}

\renewcommand{\thefigure}{\arabic{figure} (Cont.)}
\addtocounter{figure}{-1}
\begin{figure*}
\begin{center}
\begin{flushleft}

\includegraphics[scale=0.43]{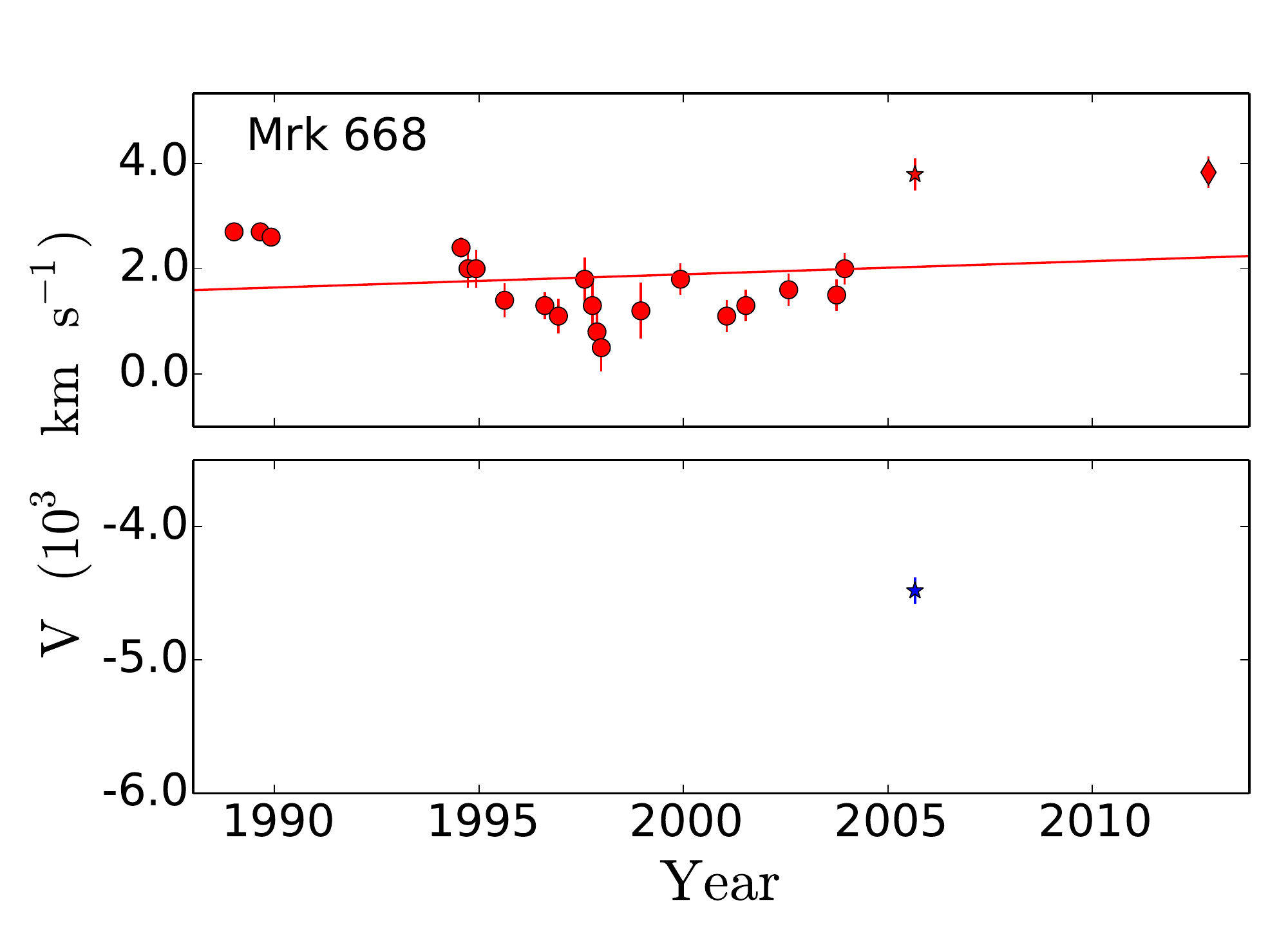}
\includegraphics[scale=0.43]{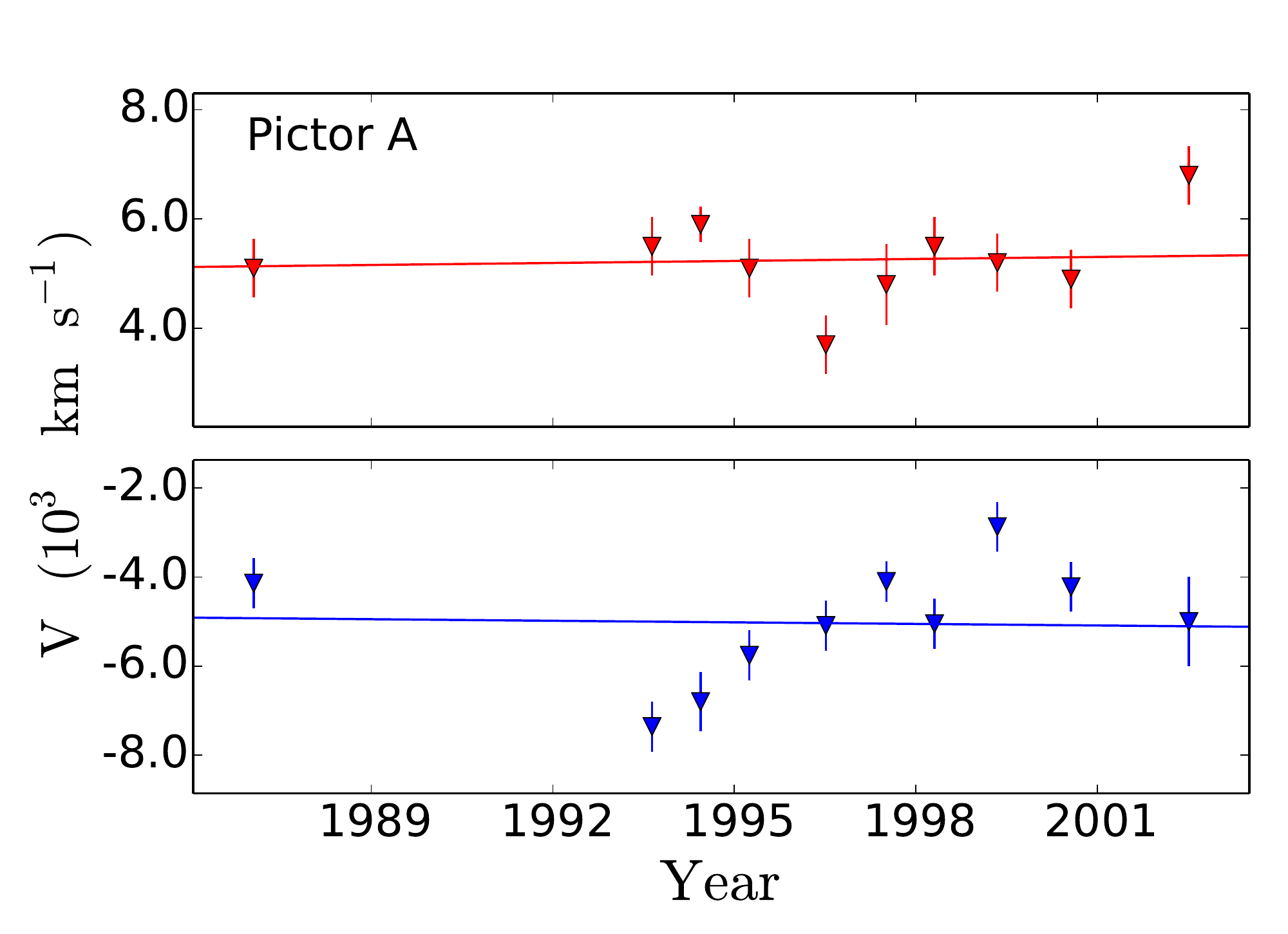}
\includegraphics[scale=0.43]{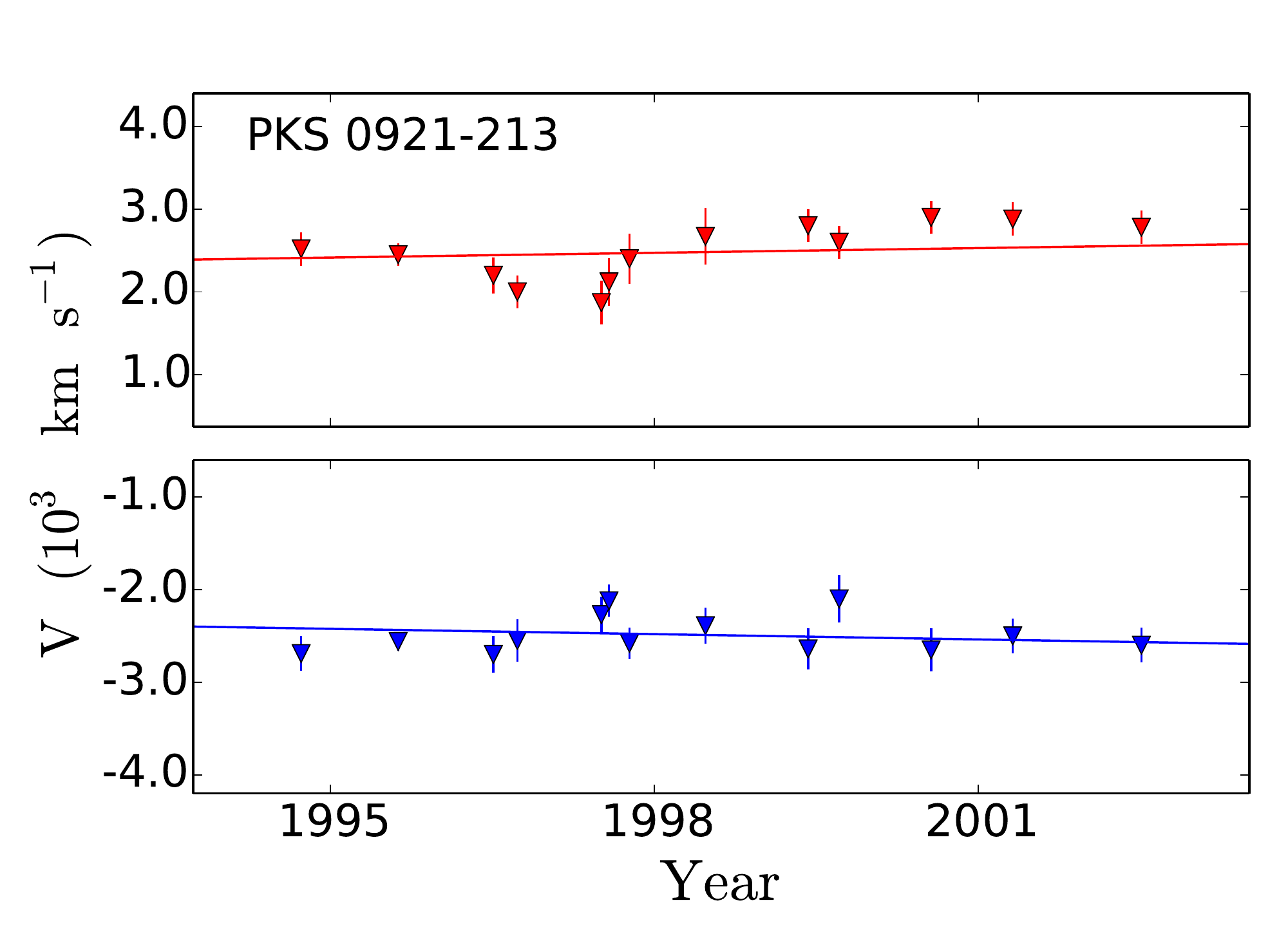}
\includegraphics[scale=0.43]{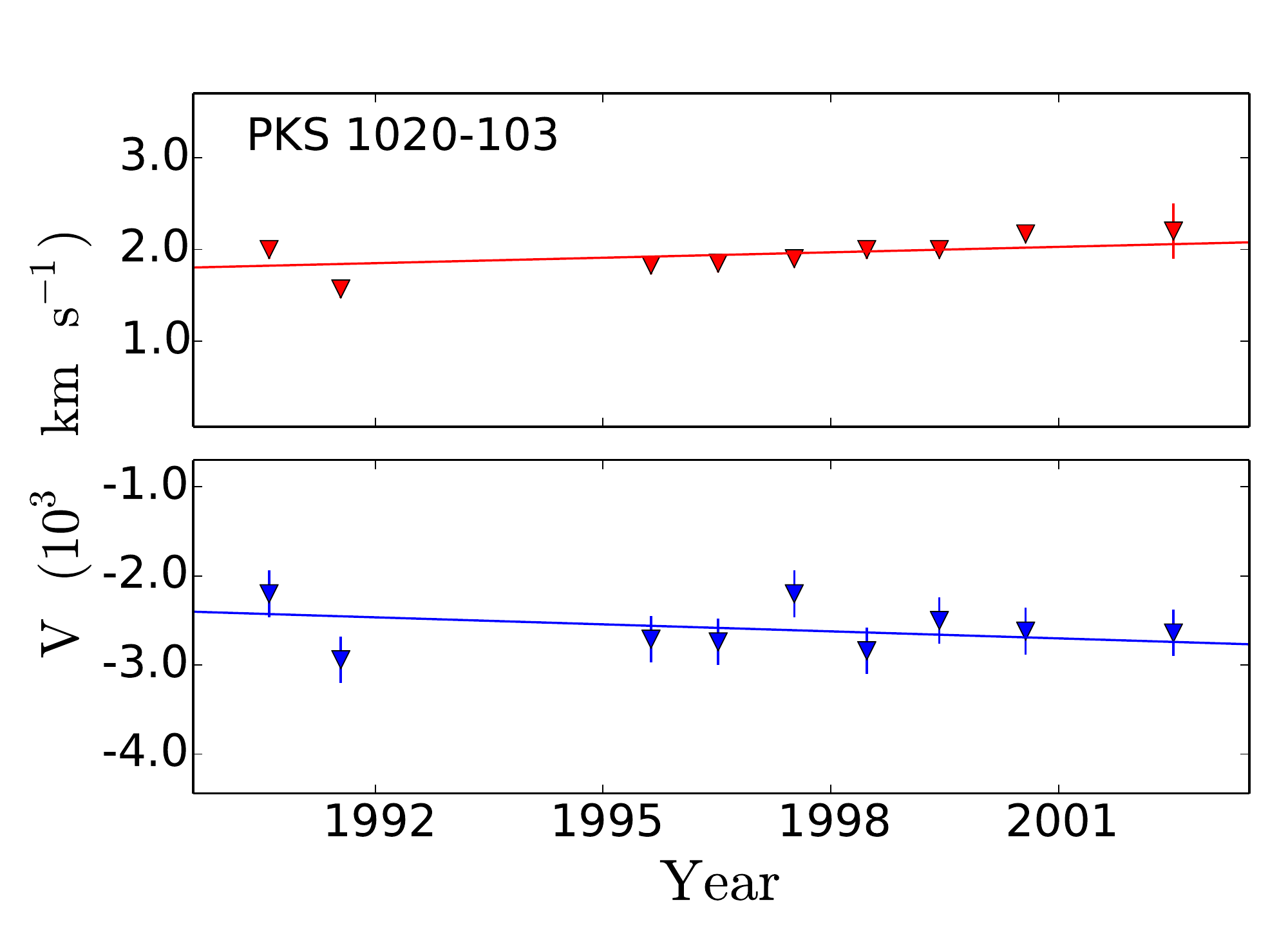}
\includegraphics[scale=0.43]{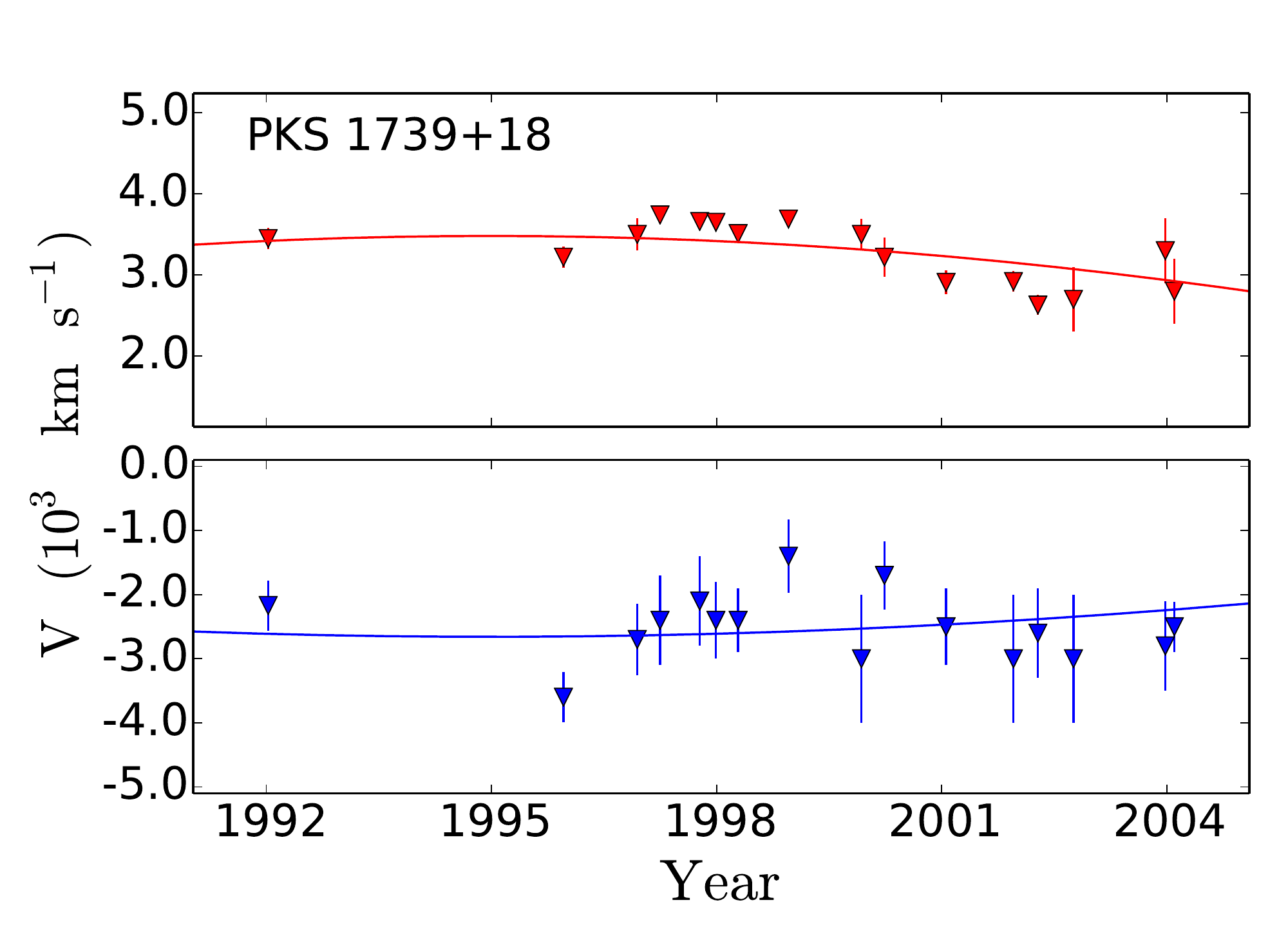}

\caption{
\label{vcurve2} Velocities of the shifted peaks of the broad H$\alpha$ line and fitted curves. 
Circle: \cite{Gezari2007};
Down-triangle: \cite{Lewis2010};
Square: \cite{Osterbrock1976};
Up-triangle: \cite{Barr1980};
Diamond: This work;
``x'': \cite{Gaskell1996};
Left-triangle: \cite{Shapovalova2013};
Right-triangle: \cite{Popovic2012};
Plus: \cite{Popovic2014};
Star: SDSS.
Solid curves are the best-fit models from the unweighted least squares method (Table~\ref{P_fitting_lsq}).
\bluetext{Note that the velocity scale on the vertical axis has been expanded to show detail (i.e., it does not start at zero).}
}
\end{flushleft}
\end{center}
\end{figure*}

Finally, we note that some objects (3C~332, 3C~59, and CBS~74) do show interesting symmetric shifts between the red and the blue Balmer line peaks. However, peaks crossing at zero velocity, a signature we expect to see in binary motion, is not observed for any of these objects. A similar behavior is seen is NGC~1097, a double-peaked emitter not included in this study \citep[see Figure~5 of][]{Schimoia2015}. In NGC~1097 the separation between the two peaks of the broad H$\alpha$ line changes in accordance with the integrated flux of the lines; as the peak separation fluctuates the two peaks never cross. A similar correlation can be discerned for 3C~390.3 by inspecting Figure 4 of \citet{Shapovalova2001}.  A short segment of the velocity curve of NGC~1097 could easily be interpreted as a segment of a sinusoid and attributed (incorrectly) to orbital motion of two BL regions.

There remains a possibility that the orbits of some SMBH binaries are eccentric, which would mean that the orbital models we have adopted for our analysis are not applicable. For a complete analysis, models of eccentric orbits should also be compared with the data. But even if such models were found to be viable and to yield reasonable BH masses, the additional arguments against the SMBH binary hypothesis presented above would remain valid.

In view of the above arguments, interpretations other than SMBH binaries should be considered  for the double peaked Balmer lines. To fit the line profiles of Arp~102B and 3C~390.3, \cite{Chen1989} proposed a thin disk illuminated by a thick hot inner torus, with only one BH present in the center.  In this model, velocity shifts can be attributed to transient bright spots or other structures in the disk \citep[e.g.,][]{Newman1997}. {Alternatively, slow, large amplitude variations in the illumination of such a disk could be responsible for the apparent radial velocity changes of the Balmer line peaks \citep[e.g.,][and references therein]{Schimoia2015}.

\section{Summary, Conclusions, and Future Prospects}\label{summary}
We searched for periodic radial velocity variations in the broad, double-peaked Balmer emission lines of 13 AGNs for which multi-decade monitoring observations are available. We obtained velocity measurements for these objects from multiple sources: \cite{Gezari2007}, \cite{Lewis2010}, \cite{Popovic2012}, \cite{Popovic2014}, follow-up observations for four AGNs at the MDM Observatory in 2009--2013, four SDSS archival spectra, and digitized graphic spectra from the literature (see Table~\ref{targetlist}). We searched for periodic signals in both flux and velocity variations, but without success. We conclude that for these AGNs, any periods are significantly longer than our monitoring span, and/or mechanisms other than orbiting BHs are responsible for their double-peaked broad H$\alpha$ lines and their line profile changes. We also presented additional arguments based on other observations that strongly disfavor the SMBH binary interpretation of broad, double-peaked emission lines. 

By extension, newer studies of single displaced-peak emission lines in AGNs will have to take into account the possibility that the same processes will manifest themselves as radial velocity variations that are not directly related to orbital motion of SMBH binaries or recoiling SMBH mergers. This  velocity ``noise'' complicates the search for kinematic evidence of SMBH motion in broad emission lines, placing more stringent observational requirements on the duration and coherence of periodic signals than have been attained so far. Potential sources of such noise include reverberation, as discussed by \citet{Barth2015}, and other phenomena that occur on longer time scales \citep[on the order of the dynamical time or longer; see illustration in Fig.~5 of][]{Schimoia2015}. It is, therefore, useful to continue monitoring the profiles of double-peaked Balmer emission lines on long time scales to check for non-monotonic changes of the velocities of the two peaks that would provide a further test of the SMBH binary hypothesis. If, for example, the trend of decreasing separation of the two peaks in some objects (e.g., 3C~332) reverses itself before the peaks cross at zero velocity (as in 3C~390.3), this would further weaken the case for SMBH binaries in double-peaked emitters.

\begin{acknowledgments}
This work was supported by grant AST-1211756 from the National Science Foundation. We acknowledge comments and suggestions from an anonymous referee.
\end{acknowledgments}

\end{document}